\documentclass[3p,,preprint,10pt]{elsarticle}
\usepackage{amssymb}
\usepackage{graphicx}
\usepackage{epstopdf}
\usepackage{epsfig}
\usepackage{times}
\usepackage{amsmath}
\usepackage{amssymb}
\usepackage{latexsym}
\usepackage{color}
\usepackage{verbatim}
\usepackage{cite}
\usepackage{BernsteinStyle2}
\usepackage{caption}
\usepackage{makecell}
\usepackage{multicol}
\usepackage{caption}
\usepackage{subcaption}
\usepackage{xcolor}
\usepackage{hyperref}
\hypersetup{
    colorlinks=true,
    linkcolor=blue,    % color of internal links
    urlcolor=blue,     % color of external links
    citecolor=blue     % color of citations
}
\newcommand{\mycite}[1]{\textcolor{black}{(\citet{#1})}}

\makeatletter
\newcommand*\bigcdot{\mathpalette\bigcdot@{.5}}
\newcommand*\bigcdot@[2]{\mathbin{\vcenter{\hbox{\scalebox{#2}{$\m@th#1\bullet$}}}}}
\makeatother
\journal{Control Engineering Practice}

\begin{document}

\begin{frontmatter}


\author[inst1]{Shashank Verma\corref{cor1}}
\ead{shaaero@umich.edu}
\cortext[cor1]{Corresponding author}
\author[inst1]{Dennis S. Bernstein}

\title{Real-Time Kinematics-Based Sensor-Fault Detection\\ for Autonomous Vehicles Using Single and Double Transport\\ with Adaptive Numerical Differentiation }


% \author[inst1]{Shashank Verma}

\affiliation[inst1]{organization={Aerospace Department, University of Michigan},%Department and Organization
            addressline={ 1320 Beal Ave}, 
            city={Ann Arbor},
            postcode={48109}, 
            state={Michigan},
            country={USA}}

%pls make the changes you suggested and then I will edit it

%we never use "our" in a paper-----it sounds too possessive

\begin{abstract}
%    
% Sensor-fault detection is crucial for ensuring the safe operation of autonomous vehicles. 
% %
% This paper introduces a novel kinematics-based approach for detecting and identifying faulty sensors, which is model-independent, rule-free, and applicable to both ground and aerial vehicles. 
% %
% Kinematics-based sensor fault detection (KSFD) relies solely on kinematic relations, sensor measurements, and real-time numerical differentiation. For ground vehicles confined to the horizontal plane, KSFD uses six kinematics-based error metrics computed in real time using onboard sensor data, including radar, rate gyro, magnetometer, and accelerometer measurements, along with their derivatives. For aerial vehicles, KSFD employs nine kinematics-based error metrics. Real-time numerical differentiation is performed using  adaptive input and state estimation (AISE)  to compute the derivatives. Simulated and experimental examples are provided to evaluate the effectiveness of KSFD.
%
Sensor-fault detection is crucial for the safe operation of autonomous vehicles. 
This paper introduces a novel kinematics-based approach for detecting and identifying faulty sensors, which is model-independent, rule-free, and applicable to ground and aerial vehicles. This method, called kinematics-based sensor fault detection (KSFD), relies on kinematic relations, sensor measurements, and real-time single and double numerical differentiation.
Using onboard data from radar, rate gyros, magnetometers, and accelerometers, KSFD uniquely identifies a single faulty sensor in real time. 
To achieve this, adaptive input and state estimation (AISE) is used for real-time single and double numerical differentiation of the sensor data, and the single and double transport theorems are used to evaluate the consistency of data. 
Unlike model-based and knowledge-based methods, KSFD relies solely on sensor signals, kinematic relations, and AISE for real-time numerical differentiation.
For ground vehicles, KSFD requires six kinematics-based error metrics, whereas, for aerial vehicles, nine error metrics are used.
Simulated and experimental examples are provided to evaluate the effectiveness of KSFD.
\end{abstract}

% let me read the whole thing

%ok, just email me later thanks.  Try not to make huge edits since that will take a lot of time for me to go over it thanks
% looks good. Thanks. 


% \begin{highlights}
% \end{highlights}


\end{frontmatter}


\section{Introduction}
\label{sec:introduction}

Sensor failure can lead to catastrophic outcomes for systems under control, since reliable sensor measurements are critical for proper system functioning. Recent incidents, such as those involving the Boeing 737 Max, attributed to a faulty angle-of-attack sensor, and the tragic crash of Air France Flight 447 on June 1, 2009, caused by a faulty airspeed sensor, highlight the tragic consequences of sensor failure in aircraft.

In autonomous vehicles, various sensors, such as cameras, radar, and LiDAR, are used to perceive the environment. Malfunctioning or failing sensors can result in severe accidents, such as collisions. For instance, a sensor failure might cause an autonomous vehicle to miss a stop sign or traffic light, miscalculate distances or velocities, or fail to detect pedestrians, objects, or other vehicles. As autonomous vehicles become more advanced, they require an increasing number of sensors, which in turn raises the risk of sensor failure. This escalating complexity necessitates advanced methods for detecting and diagnosing sensor faults.

Sensor-fault detection \mycite{SamyPostlethwaiteGu2011} is a subset of sensor diagnostics and prognostics \mycite{hwang2010survey, doraiswami, isermann1984process, isermann1997, isermann2006fault, isermann2011fault, Gertler1998, VenkatasubramanianRengaswamyYinKavuri2003, RengaswamyVenkatasubramanian2000, PattonChen1997, PattonFrankClark2000, DingFaultDiagnosis2008, FrankDing1997, ChiangRusselBraatz2001}. In some cases, sensor health can be assessed by exciting the system in a controlled manner, using a plant model and an observer to predict the response, and by comparing the measured response to the prediction. This approach, known as \textit{active} sensor-fault detection, is based on \textit{residual generation} \mycite{rajamaniIJC,frank1990fault,chow1984analytical,staroswiecki2001analytical,isermann2006fault,isermann2011fault,chen2012robust,zhangTAC,seilerADS}. In contrast, \textit{passive} sensor-fault detection detects sensor faults by analyzing each sensor signal separately and searching for anomalies \mycite{Martin1994,KothamasuHuangVerDuin2006,YanGao2007,BasirYuan2007,Basseville1988,Basseville1998,BassevilleNikiforov1993}.

Detecting sensor faults is challenging for several reasons. Sensors can fail gradually, suddenly, or intermittently, with unknown biases, scale factors, and nonlinearities. Additionally, distinguishing sensor failures from the effects of disturbances and system changes can be difficult. The validity of sensor measurements is often most questionable during rare and dangerous events when measurements are most needed to enable corrective action. Therefore, it is essential to detect and diagnose sensor faults promptly to avoid catastrophic events.

Fault detection and isolation methods can be categorized into three categories, namely, hardware redundancy, analytical redundancy, and signal processing \mycite{fourlas_survey_2021}. Hardware redundancy methods involve using multiple sensors to provide redundant measurements, increasing the system's complexity and cost \mycite{balaban_modeling_2009,blanke_diag_fault_book_2015, VenkatasubramanianRengaswamyYinKavuri2003}.

Analytical redundancy methods, including model-based and knowledge-based approaches, rely on accurate mathematical models or expert knowledge and historical data. These methods are computationally intensive and can struggle with new or unforeseen fault conditions. In autonomous vehicles, machine learning and hybrid approaches have been widely explored. A hybrid method combining the Jarque-Bera test and fuzzy systems is used by \mycite{fang_fault_2020}, while \mycite{min_fault_2023} proposes a machine learning-based anomaly detection system. Neural network-based observers for fault detection in autonomous nonlinear systems are presented in \mycite{cao_sensor_2023}. A deep neural network architecture for multi-sensor-fault detection is discussed in \mycite{safavi_multi-sensor_2021}. Model-based methods, including the quadratic Kalman filter, are used for UAV fault detection in \mycite{han_quadratic-kalman-filter-based_2022}, whereas \mycite{huang_sensor_2023} uses a Luenberger observer for sensor fault detection.

For unmanned aerial vehicles (UAVs), the survey \mycite{fourlas_survey_2021, li_recent_2020} provides an overview of various fault diagnosis and fault-tolerant control methods. Specific approaches, such as sliding mode observers for inertial measurement unit (IMU) sensor fault diagnosis in quadrotor UAVs, are detailed in \mycite{avram_imu_2015}.  General techniques have been developed for sensor-fault detection across different applications. An online fault detection and diagnosis method combining data-driven and model-based approaches is presented by \mycite{khalastchi_sensor_nodate}. Additionally, a nonlinear fault detection and diagnosis scheme for unmanned quadrotor helicopters is introduced by \mycite{zhong_sensor_2019}.

Signal-processing-based methods leverage advanced signal analysis techniques for fault detection. For aircraft, \mycite{ansari_aircraft_2016} presents a sensor fault detection method involving state and input estimation using nonlinear kinematics state space models, while \mycite{ansari_estimation_2017} focuses on estimating angular velocity and rate-gyro noise. \mycite{balaban_modeling_2009} examines the physical underpinnings of sensor faults and provides insights into fault accommodation strategies. Fault detection in aircraft using kinematic relations is explored in \mycite{chu_air_2013, van_eykeren_sensor_2014}. Additionally, \mycite{fravolini_experimental_2018} proposes data-based approaches for robust fault detection of air data sensors.

These studies underscore the importance and diversity of sensor fault detection and diagnosis methods, emphasizing the need for robust and reliable techniques to ensure the safety and reliability of autonomous systems. However, these approaches have several disadvantages. Hardware redundancy increases system cost, complexity, and maintenance due to the need for multiple sensors \mycite{balaban_modeling_2009}. Hardware redundancy also results in higher power consumption, weight, and size, which are critical constraints in applications such as autonomous vehicles and aerospace \mycite{blanke_diag_fault_book_2015}. Additionally, managing multiple sensors complicates sensor fusion algorithms, potentially causing incorrect fault diagnoses \mycite{VenkatasubramanianRengaswamyYinKavuri2003}. Furthermore, hardware redundancy may not effectively detect systematic errors that impact all sensors simultaneously, limiting its reliability \mycite{Isermann2004ModelbasedFA}.

Analytical redundancy, which includes model-based and knowledge-based sensor-fault-detection methods, also has several disadvantages. Model-based approaches require accurate mathematical models of the system, which can be difficult to develop and may not capture all real-world complexities, leading to false alarms or missed faults \mycite{Isermann2004ModelbasedFA}. These methods are computationally intensive and can be challenging to implement in real-time applications with limited processing power. Knowledge-based methods rely heavily on expert knowledge and historical data, which may not always be available or comprehensive \mycite{VenkatasubramanianRengaswamyYinKavuri2003}. Furthermore, these methods can struggle with new or unforeseen fault conditions that were not considered during the development phase, limiting their adaptability and robustness \mycite{blanke_diag_fault_book_2015}. Both approaches can also suffer from high implementation and maintenance costs, as they require continuous updates and validation to remain effective \mycite{Isermann2004ModelbasedFA}.

The contribution of this paper is the development and demonstration, both numerically and experimentally, of a novel, real-time-implementable, kinematics-based sensor fault detection (KSFD) approach. KSFD, classified as a signal processing-based method, detects and uniquely identifies the occurrence of a single faulty sensor.
KSFD uses kinematics-based error metrics derived from the single- and double-transport theorems, along with real-time numerical differentiation for sensor fault detection. 
Real-time single and double numerical differentiation are performed using adaptive input and state estimation (AISE), as developed in \mycite{verma_shashank_2023_realtime_IJC, verma_shashank_2024_realtime_VRF_axiv}. 
AISE is chosen for numerical differentiation for KSFD because it estimates the derivative causally in real time using sampled data while adapting to unknown and changing sensor-noise characteristics. 
AISE facilitates the practical implementation of KSFD.

Unlike model-based and knowledge-based methods, KSFD relies solely on sensor signals, kinematic relations, and AISE for real-time numerical differentiation.
The present paper thus represents the first use of single and double transport along with single and double real-time numerical differentiation to detect faulty sensors.
The authors are not aware of any other sensor-fault-detection technique that operates under these kinematics-based assumptions.

As developed and demonstrated in the present paper, KSFD detects and identifies the occurrence of a single faulty sensor. 
In cases where two or more sensors fail simultaneously, KSFD indicates the presence of at least one faulty sensor but does not determine which sensors are faulty.
However, the scope of this paper is limited to the case of a single sensor fault.
%
%new sentence
%
%NOT for the arxiv update
%
% A preprint of the present paper is available as arXiv:2309.05158.

%
%
%
The contents of this paper are organized as follows. Section \ref{sec:problem_formulation} presents the single and double transport theorems. Section \ref{sec:data_single_double_transport} discusses the use of sensor data within these theorems. Section \ref{sec:AISE} describes AISE. Section \ref{sec:sensor_fault_Detection} introduces the error metrics for sensor-fault detection and their application to ground and aerial vehicles. Section \ref{sec:num_example_simu} provides a simulated example of sensor-fault detection in an accelerometer. 
Section \ref{sec:num_example_exp_2d} presents two experimental examples using real-world data, illustrating KSFD for radar and rate-gyro failures in a ground vehicle. 
Finally, Section \ref{sec:num_example_exp_3d} provides an experimental investigation of KSFD for accelerometer failure in an aerial vehicle.

%%%%%%%%%%%%%%%%%%%%%%%%%%%%%%%%%%%%%%%%%%%%%%%%%%%%%%%%%%%
\section{Problem Formulation}
\label{sec:problem_formulation}

We assume the Earth is inertially nonrotating and nonaccelerating.
The right-handed frame $\rmF_{\rm E} = \begin{bmatrix} \hat{\imath}_{\rmE} & \hat{\jmath}_{\rmE} & \hat{k}_{\rmE} \end{bmatrix}$ is fixed to the Earth, and the origin $\rmo_{\rmE}$ of $\rmF_{\rmE}$ is any convenient point fixed on the Earth;
hence, $\rmo_{\rmE}$ has zero inertial acceleration.
$ \hat{k}_{\rmE}$ points downward, and $\hat{\imath}_{\rmE}$ and $\hat{\jmath}_{\rmE}$ are horizontal. 
The right-handed vehicle body-fixed frame is denoted by $\rmF_{\rm B} = \begin{bmatrix} \hat{\imath}_{\rmB} & \hat{\jmath}_{\rmB} & \hat{k}_{\rmB} \end{bmatrix}$.  
The origin $\rmo_{\rmB}$ of $\rmF_{\rmB}$ is 
any point fixed on the vehicle, 
$\hat{\imath}_{\rmB}$ is pointing forward of the vehicle, $\hat{\jmath}_{\rmB}$ is directed out the right side of the vehicle, and $\hat{k}_{\rmB}$ is directed downward.

Next, $\rmF_{\rmE}$ and $\rmF_{\rmB}$ are related by
\begin{equation}
    \rmF_{\rmB} = \ \tarrow{R}_{\rm B/E}  \rmF_{\rmE},
\end{equation}
where $ \tarrow{R}_{\rm B/E}$ is the physical rotation matrix represented by a 3-2-1 azimuth-elevation-bank Euler rotation sequence involving two intermediate frames $F_{\rmE'}$ and $F_{\rmE''}$. In particular,
\begin{equation}
    \tarrow{R}_{\rm B/E} \ = \ \tarrow{R}_{\hat{\imath}_{\rmE''}}(\Phi) \tarrow{R}_{\hat{\jmath}_{\rmE'}}(\Theta) \tarrow{R}_{\hat{k}_{\rmE}}(\Psi), \label{rotation_mat}
\end{equation}
where $\rmF_{\rmE'} =  \tarrow{R}_{\rm E'/E} \rmF_{\rmE}$, $\rmF_{\rmE''} =  \tarrow{R}_{\rm E''/E'} \rmF_{\rmE'}$, the Euler angles $\Psi, \Theta, \Phi \in (-\pi, \pi] $, and the Rodrigues rotation matrix is given by
\begin{equation}
     \tarrow{R}_{\hat{n}}(\kappa) \isdef (\cos{\kappa})  \tarrow{I} + (1- \cos{\kappa}){\hat{n}}{\hat{n}}^{'} + (\sin{\kappa}){\hat{n}}^{\times}, \label{rodrigues}
\end{equation} 
where $\tarrow{I}$ is the physical identity matrix, and the superscript $\times$ creates a skew-symmetric physical matrix. Note that \eqref{rodrigues} represents a right-hand-rule rotation about the eigenaxis $\hat{n}$ by the eigenangle $\kappa$ according to the right-hand rule.
The physical angular velocity $\vect{\omega}_{\rmB/\rmE}$ of $\rmF_{\rmB}$ relative to $\rmF_{\rmE}$ is defined by Poisson's equation 
\begin{equation}
      \framedotB{\tarrow{R}}_{\rm B/E} \ = \ \tarrow{R}_{\rm B/E} \vect{\omega}_{\rmB/\rmE}^{\times},
\end{equation}
where $\framedotB{}$ denotes the time derivative with respect to $\rmF_{\rmB}$.

The location of the vehicle center of mass $\rmo_{\rmB}$ relative to $\rmo_{\rmE}$ at each time instant is given by the physical position vector $\vect{r}_{\rmo_{\rmB}/\rmo_{\rmE}}$.
The first and second derivatives of $\vect{r}_{\rmo_{\rmB}/\rmo_{\rmE}}$ with respect to $\rmF_{\rmE}$  are given by the single and double transport theorems, respectively, which have the forms
\begin{equation}
\begin{aligned}
   \framedotE{\vect{r}}_{\rmo_{\rmB}/\rmo_{\rmE}} &=\, \framedotB{\vect{r}}_{\rmo_{\rmB}/\rmo_{\rmE}} +\, \vect{\omega}_{\rmB/\rmE} \times \vect{r}_{\rmo_{\rmB}/\rmo_{\rmE}}, \end{aligned} \label{single_transport}
\end{equation}
\begin{equation}
\begin{aligned}
   \frameddotE{\vect{r}}_{\rmo_{\rmB}/\rmo_{\rmE}} &=\, \frameddotB{\vect{r}}_{\rmo_{\rmB}/\rmo_{\rmE}} +\, 
    2\vect{\omega}_{\rmB/\rmE}\, \times \framedotB{\vect{r}}_{\rmo_{\rmB}/\rmo_{\rmE}} + \framedotB{\vect{\omega}}_{\rmB/\rmE} \times \vect{r}_{\rmo_{\rmB}/\rmo_{\rmE}}  + \vect{\omega}_{\rmB/\rmE} \times (\vect{\omega}_{\rmB/\rmE} \times \vect{r}_{\rmo_{\rmB}/\rmo_{\rmE}}).
\end{aligned} \label{double_transport}
\end{equation}
Note that \eqref{single_transport} and \eqref{double_transport} are exact kinematic relations.

%%%%%%%%%%%%%%%%%%%%%%%%%%%%%%%%%%%%%%%%%%%%%%%%%%%%%%%%%%%
\section{Using Sensor Data in Single and Double Transport} \label{sec:data_single_double_transport}

The vehicle is equipped with onboard sensors, including magnetometer, radar, rate gyros, and accelerometers.
Some radar sensors use the Doppler effect to estimate velocity, in this article we assume that the radar  provides only position data. 
To evaluate the terms in \eqref{single_transport} and \eqref{double_transport} using sensor data, we begin by resolving all the terms on the right-hand side (RHS) of \eqref{single_transport} and \eqref{double_transport} in $\rmF_{\rm B}$, which yields
\begin{align}
    \begin{bmatrix}
    r_{x} \\ r_{y} \\ r_{z} 
    \end{bmatrix} \isdef \vect{r}_{\rmo_{\rmB}/\rmo_{\rmE}}\bigg\vert_\rmB, \quad  
\begin{bmatrix}
    \dot{r}_{x} \\ \dot{r}_{y} \\ \dot{r}_{z} 
\end{bmatrix}=\,\framedotB{\vect{r}}_{\rmo_{\rmB}/\rmo_{\rmE}}\bigg\vert_\rmB, \quad 
%\label{vel_pos_resolve_B}
%\end{align}
%
%\begin{align}
    \begin{bmatrix}
    \ddot{r}_{x} \\ \ddot{r}_{y} \\ \ddot{r}_{z} 
    \end{bmatrix} = \frameddotB{\vect{r}}_{\rmo_{\rmB}/\rmo_{\rmE}}\bigg\vert_\rmB  , \quad \label{acc_vel_resolve_B}
\end{align}
\begin{align}
    \begin{bmatrix}
    \omega_{x} \\ \omega_{y} \\ \omega_{z} 
    \end{bmatrix} \isdef \vect{\omega}_{\rmB/\rmE}\bigg\vert_\rmB, \quad  
    \begin{bmatrix}
    \dot{\omega}_{x} \\ \dot{\omega}_{y} \\ \dot{\omega}_{z} 
    \end{bmatrix} =\,  \framedotB{\vect{\omega}}_{\rmB/\rmE}\bigg\vert_\rmB. \label{angular_vel_acc_resolve_B}
\end{align}
We resolve \eqref{rotation_mat} in $\rmF_{\rm B}$ to obtain the orientation  (direction cosine) matrix $\mathcal{O}_{\rmE/\rmB}$ given by
% \begin{subequations}
\begin{align}
    \mathcal{O}_{\rmE/\rmB}&\isdef \mathcal{O}_{\rmE/\rmB}(\Phi, \Theta, \Psi) \isdef {\tarrow{R}}_{\rm B/E}\bigg\vert_\rmB  
    = \ \tarrow{R}_{\hat{\imath}_{\rmE''}}(\Phi)\bigg\vert_\rmB \tarrow{R}_{\hat{\jmath}_{\rmE'}}(\Theta)\bigg\vert_\rmB \tarrow{R}_{\hat{k}_{\rmE}}(\Psi)\bigg\vert_\rmB \nonumber\\
     &= \scriptsize\begin{bmatrix}
        (\cos{\Theta})\cos{\Psi} & (\cos{\Theta})\sin{\Psi} & -\sin{\Theta} \\
        (\sin{\Phi})(\sin{\Theta})\cos{\Psi} - (\cos{\Phi})\sin{\Psi} & (\sin{\Phi})(\sin{\Theta})\sin{\Psi} + (\cos{\Phi})\cos{\Psi} & (\sin{\Phi})\cos{\Theta} \\
        (\cos{\Phi})(\sin{\Theta})\cos{\Psi} + (\sin{\Phi})\sin{\Psi} & (\cos{\Phi})(\sin{\Theta})\sin{\Psi} - (\sin{\Phi})\cos{\Psi} & (\cos{\Phi})\cos{\Theta}
    \end{bmatrix}^{\rmT}.  
\end{align}

To resolve the left-hand side (LHS) of \eqref{single_transport} in $\rmF_{\rm B}$, we first resolve the position vector $\vect{r}_{\rmo_{\rmB}/\rmo_{\rmE}}$ in $\rmF_{\rm E}$ to obtain
\begin{align}
    \begin{bmatrix}
    R_{x} \\ R_{y} \\ R_{z} 
    \end{bmatrix} \isdef \vect{r}_{\rmo_{\rmB}/\rmo_{\rmE}}\bigg\vert_\rmE = \mathcal{O}_{\rmE/\rmB}\begin{bmatrix}
    r_{x} \\ r_{y} \\ r_{z} 
    \end{bmatrix}, \label{Rxyz} 
\end{align}
and then we differentiate \eqref{Rxyz} once and twice to obtain
\begin{align}
    \begin{bmatrix}
    \dot{R}_{x} \\ \dot{R}_{y} \\ \dot{R}_{z} 
    \end{bmatrix}  =\, \framedotE{\vect{r}}_{\rmo_{\rmB}/\rmo_{\rmE}}\bigg\vert_\rmE = \frac{\rmd}{\rmd t} \left( \mathcal{O}_{\rmE/\rmB}\begin{bmatrix}
    r_{x} \\ r_{y} \\ r_{z} 
    \end{bmatrix}\right), \label{vel_resolve_E_diff} 
\end{align}
\vspace{-.15in}
\begin{align}
    \begin{bmatrix}
    \ddot{R}_{x} \\ \ddot{R}_{y} \\ \ddot{R}_{z} 
    \end{bmatrix}  =\, \frameddotE{\vect{r}}_{\rmo_{\rmB}/\rmo_{\rmE}}\bigg\vert_\rmE = \frac{\rmd^2}{\rmd t^2} \left( \mathcal{O}_{\rmE/\rmB} \begin{bmatrix}
    r_{x} \\ r_{y} \\ r_{z} 
    \end{bmatrix}\right). \label{acc_resolve_E_diff} 
\end{align}
Now, define  
\begin{align}
    \begin{bmatrix}
    V_{x} \\ V_{y} \\ V_{z} 
    \end{bmatrix} \isdef\, \framedotE{\vect{r}}_{\rmo_{\rmB}/\rmo_{\rmE}}\bigg\vert_\rmB = \mathcal{O}_{\rmB/\rmE}\begin{bmatrix}
    \dot{R}_{x} \\ \dot{R}_{y} \\ \dot{R}_{z} 
    \end{bmatrix}, \label{vel_dotE_resolve_B}
\end{align}
\vspace{-.15in}
\begin{align}
    \begin{bmatrix}
    A_{x} \\ A_{y} \\ A_{z} 
    \end{bmatrix} \isdef\, \frameddotE{\vect{r}}_{\rmo_{\rmB}/\rmo_{\rmE}}\bigg\vert_\rmB = \mathcal{O}_{\rmB/\rmE}\begin{bmatrix}
    \ddot{R}_{x} \\ \ddot{R}_{y} \\ \ddot{R}_{z} 
    \end{bmatrix}, \label{acc_dotE_resolve_B}
\end{align}
where $\mathcal{O}_{\rmB/\rmE} = \mathcal{O}^{\rmT}_{\rmE/\rmB}. $ Resolving \eqref{single_transport} and \eqref{double_transport} in $\rmF_{\rmB}$, and rewriting \eqref{acc_dotE_resolve_B} yields
\begin{align}
   \begin{bmatrix}
    V_{x} \\ V_{y} \\ V_{z} 
    \end{bmatrix} &= \begin{bmatrix}
    \dot{r}_{x} \\ \dot{r}_{y} \\ \dot{r}_{z} 
    \end{bmatrix} +  \begin{bmatrix}
    \omega_{x} \\ \omega_{y} \\ \omega_{z} 
    \end{bmatrix} \times \begin{bmatrix}
    r_{x} \\ r_{y} \\ r_{z} 
    \end{bmatrix} ,\label{single_transport_resolve_B}
    \end{align} 
\begin{align}
    \begin{bmatrix}
    A_{x} \\ A_{y} \\ A_{z} 
    \end{bmatrix} &= \begin{bmatrix}
    \ddot{r}_{x} \\ \ddot{r}_{y} \\ \ddot{r}_{z} 
    \end{bmatrix} + 
    2 \begin{bmatrix}
    \omega_{x} \\ \omega_{y} \\ \omega_{z} 
    \end{bmatrix} \times \begin{bmatrix}
    \dot{r}_{x} \\ \dot{r}_{y} \\ \dot{r}_{z} 
    \end{bmatrix} +  \begin{bmatrix}
    \dot{\omega}_{x} \\ \dot{\omega}_{y} \\ \dot{\omega}_{z} 
    \end{bmatrix} \times \begin{bmatrix}
    r_{x} \\ r_{y} \\ r_{z} 
    \end{bmatrix} +  \begin{bmatrix}
    \omega_{x} \\ \omega_{y} \\ \omega_{z} 
    \end{bmatrix} \times \left( \begin{bmatrix}
    \omega_{x} \\ \omega_{y} \\ \omega_{z} 
    \end{bmatrix} \times \begin{bmatrix}
    r_{x} \\ r_{y} \\ r_{z} 
    \end{bmatrix}\right), \label{double_transport_resolve_B}
\end{align} 
\begin{equation}
\begin{aligned}
     \begin{bmatrix}
    A_{x} \\ A_{y} \\ A_{z} 
    \end{bmatrix} = \mathcal{O}_{\rmB/\rmE}\begin{bmatrix}
    \ddot{R}_{x} \\ \ddot{R}_{y} \\ \ddot{R}_{z} 
    \end{bmatrix}. 
    \end{aligned}\label{acc_ddotE}
\end{equation}
Furthermore, \eqref{single_transport_resolve_B} and \eqref{double_transport_resolve_B} are rewritten as
\begin{align}
   \begin{bmatrix}
    V_{x} \\ V_{y} \\ V_{z} 
    \end{bmatrix} &= \begin{bmatrix}
    \dot{r}_{x} + (\omega_y r_z - \omega_z r_y) \\ \dot{r}_{y} + (\omega_z r_x - \omega_x r_z)\\ \dot{r}_{z} + (\omega_x r_y - \omega_y r_x)  
    \end{bmatrix}, \label{single_transport_resolve_B_long}
    \end{align}
\begin{align}
    \begin{bmatrix}
    A_{x} \\ A_{y} \\ A_{z} 
    \end{bmatrix} &= \begin{bmatrix}
    \ddot{r}_x + 2(\dot{r}_z\omega_y - \dot{r}_y\omega_z)+(r_z\dot{\omega}_y - r_y\dot{\omega}_z) + \omega_y(r_y\omega_x - r_x\omega_y) + \omega_z(r_z\omega_x - r_x\omega_z) \\
    \ddot{r}_y + 2(\dot{r}_x\omega_z - \dot{r}_z\omega_x)+(r_x\dot{\omega}_z - r_z\dot{\omega}_x) + \omega_z(r_z\omega_y - r_y\omega_z) - \omega_x(r_y\omega_x - r_x\omega_y) \\
    \ddot{r}_z + 2(\dot{r}_y\omega_x - \dot{r}_x\omega_y)+(r_y\dot{\omega}_x - r_x\dot{\omega}_y) - \omega_x(r_z\omega_x - r_x\omega_z) -\omega_y(r_z\omega_y - r_y\omega_z) 
    \end{bmatrix}.
\label{double_transport_resolve_B_long}
    \end{align}

\subsection{Ground Vehicles} \label{sub:ground_veh}

\begin{figure}[h!t]
              \begin{center}
            {\includegraphics[width=0.6\linewidth]{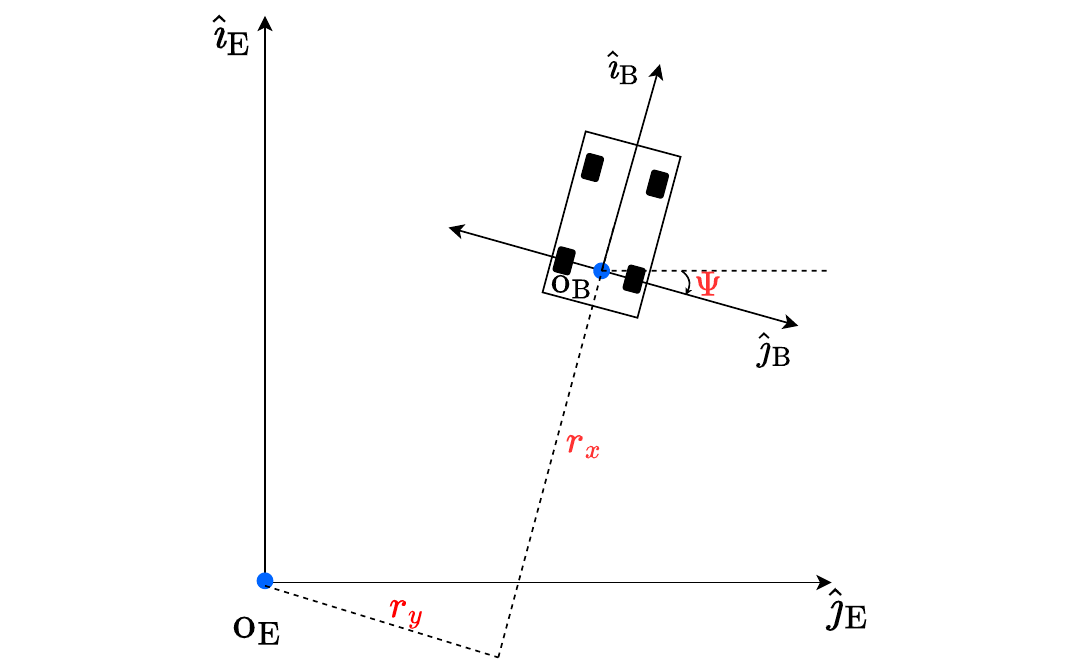}}
            \end{center}
            \caption{ Coordinate frames for the vehicle kinematics under the assumption that the vehicle remains horizontal. $\hat{\imath}_\rmE$ points north, while $\hat{\jmath}_\rmE$ points east. The vectors $\hat{k}_{\rmE}$ and $\hat{k}_{\rmB}$ point vertically downward. $r_{x}, r_{y},$ and $\Psi$ are radar and magnetometer measurements, respectively, as summarized in Table \ref{Tab:SensorData_ground}.} 
            \label{fig:diag_vehicle_coordinate_frame}
\end{figure}

For ground vehicles, we assume that the azimuth $\Psi$ of the vehicle may vary with time, but otherwise, the vehicle remains horizontal, and thus the elevation $\Theta$ and bank $\Phi$ angles are identically zero.
Table \ref{Tab:SensorData_ground} shows the data available from onboard sensors along with the derivatives that are computed to evaluate all of the terms in \eqref{single_transport_resolve_B}, \eqref{double_transport_resolve_B}, and \eqref{acc_ddotE}. To compute these derivatives, we apply AISE described in Section \ref{sec:AISE}. Figure \ref{fig:diag_vehicle_coordinate_frame} shows the coordinate system for ground-vehicle kinematics. Note that, for ground vehicles, we ignore the $z$ component of the radar and accelerometer data as well as the $x$ and $y$ components of the rate gyro.

\begin{center} 
\begin{tabular}{ |c|c|c|c| }%{ |c{2.5cm}||c{2.5cm}|c{2.5cm}|}
 \hline
 \textbf{Sensors} & \textbf{Data} & \thead{Processed \\ Data}   \\
 \hline
 Radar & \makecell{$r_{x},r_{y}$ }  & \makecell{$\dot{r}_{x}, \dot{r}_{y}$, \\ $\ddot{r}_{x}, \ddot{r}_{y}$}  \\
 \hline
  Accelerometer & $A_{x},A_{y}$ & -  \\
 \hline
 Magnetometer & $\mathcal{O}_{\rmB/\rmE} (0, 0, \Psi)$ &  -   \\
 \hline
 Rate gyro & $\omega_{z}$ &  $\dot{\omega}_{z}$   \\
 \hline

\end{tabular}
\captionof{table}{On-board sensors for ground vehicles.  These sensors, along with numerical differentiation, provide the data needed to evaluate all of the terms in \eqref{single_transport_resolve_B}, \eqref{double_transport_resolve_B}, and \eqref{acc_ddotE}.
}\label{Tab:SensorData_ground}
\end{center}

\subsection{Aerial Vehicles}  \label{sub:aerial_veh}

For aerial vehicles, the 3-2-1 Euler angles $\Psi$, $\Theta$, and $\Phi$ represent
azimuth, elevation, and bank, respectively.
In addition, the components $\omega_x,$ $\omega_y,$ and $\omega_z$ of the angular velocity vector ${\vect{\omega}}_{\rmB/\rmE}$ resolved in $\rmF_\rmB$ as defined  in \eqref{angular_vel_acc_resolve_B} are the roll rate, pitch rate, and yaw rate, respectively.
Table \ref{Tab:SensorData_aerial} lists the data available from onboard sensors as well as the computed derivatives needed to evaluate all of the terms in \eqref{single_transport_resolve_B}, \eqref{double_transport_resolve_B}, and \eqref{acc_ddotE}. These derivatives are computed using AISE given in Section \ref{sec:AISE}.

\begin{center} 
\begin{tabular}{ |c|c|c|c| }%{ |c{2.5cm}||c{2.5cm}|c{2.5cm}|}
 \hline
 \textbf{Sensors} & \textbf{Data} & \thead{Processed \\ Data}   \\
 \hline
 Radar & \makecell{$r_{x},r_{y},r_{z}$ }  & \makecell{$\dot{r}_{x}, \dot{r}_{y}, \dot{r}_{z}$, \\ $\ddot{r}_{x}, \ddot{r}_{y}, \ddot{r}_{z}$}  \\
 \hline
 Accelerometer & $A_{x},A_{y},A_{z}$ &  - \\
 \hline
 IMU & $\mathcal{O}_{\rmB/\rmE}(\Phi, \Theta, \Psi)$ &   -  \\
 \hline
 Rate gyro & $\omega_{x},\omega_{y},\omega_{z}$ &  $\dot{\omega}_{x},\dot{\omega}_{y},\dot{\omega}_{z}$   \\
  \hline
\end{tabular}
\captionof{table}{On-board sensors for aerial vehicles.  These sensors, along with numerical differentiation, provide the data needed to evaluate all of the terms in \eqref{single_transport_resolve_B}, \eqref{double_transport_resolve_B}, and \eqref{acc_ddotE}.
}\label{Tab:SensorData_aerial}
\end{center}
%
          
%%%%%%%%%%%%%%%%%%%%%%%%%%%%%%%%%%%%%%%%%%%%%%%%%%%%%%%%%%%
\section{Adaptive Input and State Estimation}  \label{sec:AISE}
We summarize AISE \mycite{verma_shashank_2023_realtime_IJC,verma_shashank_2024_realtime_VRF_axiv,verma_shashank_ACC2022} as it is applied to real-time numerical differentiation for computing the processed data listed in Table \ref{Tab:SensorData_ground} and Table \ref{Tab:SensorData_aerial}.

Consider the linear discrete-time SISO system
\vspace{-0.75em}
\begin{align}
	x_{k+1} &=  A x_{k} + Bd_{k}, 	\label{state_eqn}\\
	y_k  &= C x_k + D_{2,k} v_k, \label{output_eqn}
\end{align}
where
$k\ge0$ is the step,
$x_k \in \mathbb R^{n}$ is the unknown state,
$d_k \in \mathbb R$ is unknown input,
$y_k \in \mathbb R$ is a measured output,
$v_k \in \mathbb R$ is standard white noise, 
and $D_{2,k}v_k \in \mathbb R$ is the sensor noise at time $t = kT_\rms$, where $T_\rms$ is the sample time.
The matrices $A \in \mathbb R^{n \times n}$, $B \in \mathbb R^{n \times 1}$, and $C \in \mathbb R^{1 \times n}$ are assumed to be known, and $D_{2,k}$ is assumed to be unknown.
The sensor-noise covariance is $V_{2,k} \isdef D_{2,k} D_{2,k}^\rmT$.
The goal of adaptive input estimation (AIE) is to estimate $d_k$ and $x_k$.

In the application of AIE to real-time numerical differentiation, we use \eqref{state_eqn} and \eqref{output_eqn} to model a discrete-time integrator. As a result, AIE furnishes an estimate denoted by $\hat{d}_k$ for the derivative of the sampled output $y_k$. 
For single discrete-time differentiation,  $A = 1, B = T_\rms,$ and $C=1$, whereas, 
for double discrete-time differentiation,  
\begin{align}
    A = \begin{bmatrix}
        1 & T_\rms\\ 0 & 1
    \end{bmatrix}, \quad B = \begin{bmatrix}
       \half T^2_\rms \\ {T_\rms}
    \end{bmatrix}, \quad C = \begin{bmatrix}
        1 & 0
    \end{bmatrix}.
\end{align}

\vspace{-0.5em}

\subsection{Input Estimation}
AIE comprises three subsystems, namely, the Kalman filter forecast subsystem, the input-estimation subsystem, and the Kalman filter data-assimilation subsystem.
First, consider the Kalman filter forecast step
%\vspace{-0.75em}
%
\begin{gather}
	x_{{\rm fc},k+1} = A x_{{\rm da},k} + B \hat{d}_{k},	\label{kalman_fc_state}\\
	y_{{\rm fc},k} =  C x_{{\rm fc},k}, \label{kalman_fc_output}\\
	z_k = y_{{\rm fc},k} - y_k, 		\label{innov_error}
\end{gather}
where
$x_{\rm da,k} \in \mathbb R^{n}$ is the data-assimilation state, 
$x_{{\rm fc},k} \in \mathbb R^{n}$ is the forecast state,
$\hat d_k$ is the estimate of $d_k$, 
$y_{\rmf\rmc,k} \in \mathbb R$ is the forecast output,
$z_k \in \mathbb R$ is the residual, and $x_{{\rm fc},0} = 0$.

Next, to obtain $\hat{d}_k$, the input-estimation subsystem of order $n_\rme$ is given by the exactly proper, input-output dynamics
\vspace{-0.5em}
\begin{align}
\hat{d}_k = \sum\limits_{i=1}^{n_\rme} P_{i,k} \hat{d}_{k-i} + \sum\limits_{i=0}^{n_\rme} Q_{i,k} z_{k-i}, \label{estimate_law1}
\end{align}
%
% %
where $P_{i,k} \in \BBR$ and $Q_{i,k} \in \BBR$.
AIE minimizes a cost function that depends on $z_{k}$ by updating $P_{i,k}$ and $Q_{i,k}$ as shown below.
The subsystem \eqref{estimate_law1} can be reformulated as
\vspace{-0.75em}
\begin{align}
\hat{d}_k=\Phi_k \theta_k, \label{estimate_law12}
\end{align}
where the estimated coefficient vector $\theta_k \in \mathbb{R}^{l_{\theta}}$ is defined by
\vspace{-0.75em}
\begin{align}
\hspace{-0.2cm}\theta_k \isdef \begin{bmatrix}
P_{1,k} & \cdots & P_{n_{\rme},k} & Q_{0,k} & \cdots & Q_{n_{\rme},k}
\end{bmatrix}^{\rmT}, \label{est_coeff_vec}
\end{align}
the regressor matrix $\Phi_k \in \mathbb{R}^{1 \times l_{\theta}}$ is defined by
\vspace{-0.5em}
\begin{align}
	\hspace{-0.2cm}\Phi_k \isdef
		\begin{bmatrix}
			\hat{d}_{k-1} &
			\cdots &
			\hat{d}_{k-n_{\rme}} &
			z_k &
			\cdots &
			z_{k-n_{\rme}}
		\end{bmatrix},
\end{align}
and $l_\theta \isdef  2n_{\rme} +1$.
The subsystem \eqref{estimate_law1} can be written using backward shift operator $\bfq^{-1}$ as
\vspace{-0.25em}
\begin{align}
   \hat{d}_{k} = G_{\hat{d}z,k}(\bfq^{-1})z_k,
\end{align}
where
\vspace{-0.25em}
\begin{align}
    G_{\hat{d}z,k} &\isdef D_{\hat{d}z, k}^{-1}  \it{N}_{\hat{d}z,k}, \label{d_hat_z_tf} \\
    D_{\hat{d}z,k}(\bfq^{-1}) &\isdef I_{l_d}-P_{1,k}\bfq^{-1} - \cdots-P_{n_\rme,k}\bfq^{-n_\rme}, \label{d_hat_z_tf_D} \\
    N_{\hat{d}z, k}(\bfq^{-1}) &\isdef Q_{0,k} + Q_{1,k} \bfq^{-1}+\cdots+Q_{n_\rme,k}\bfq^{-n_\rme}. \label{d_hat_z_tf_N}
\end{align}
Next, define the filtered signals
\vspace{-0.5em}
\begin{align}
\Phi_{{\rm f},k} &\isdef G_{{\rm f}, k}(\bfq^{-1}) \Phi_{k}, \quad
\hat{d}_{{\rm f},k} \isdef G_{{\rm f}, k}(\bfq^{-1}) \hat{d}_{k}, \label{eq:filtdhat}
\end{align}
where, for all $k\ge 0$,
\vspace{-0.75em}
\begin{align}
G_{{\rm f}, k}(\bfq^{-1}) = \sum\limits_{i=1}^{n_{\rm f}} \bfq^{-i}H_{i,k}, \label{Gf}
\end{align}
\vspace{-1em}
\begin{align}
H_{i,k} &\isdef \left\{
\begin{array}{ll}
C B, & k\ge i=1,\\
C \overline{A}_{k-1}\cdots \overline{A}_{k-(i-1)}  B, & k\ge i \ge 2, \\
 % C \left(\prod_{j=1}^{i-1} \overline{A}_{k-j} \right) B, & k\ge i \ge 2, \\
0, & i>k,
\end{array}
\right. 
\end{align}
and $\overline{A}_k \isdef A(I + K_{{\rm da},k}C)$, where $K_{{\rm da},k}$ is the Kalman filter gain given by \eqref{kalman_gain} below.
Furthermore, for all $k \ge 0$, define the {\it retrospective performance variable} $z_{{\rm r},k} \colon \BBR^{l_\theta} \rightarrow \BBR$ by
\vspace{-0.75em}
\begin{align}
z_{{\rm r},k}(\hat{\theta}) \isdef z_k -( \hat{d}_{{\rm f},k} - \Phi_{{\rm f},k}\hat{\theta} ), \label{eq:RetrPerfVar} 
\end{align}
and define the \textit{retrospective cost function} $\SJ_k \colon \BBR^{l_\theta} \rightarrow \BBR$ by
\vspace{-0.75em}
\begin{align}
    \SJ_k(\hat{\theta}) \isdef  \sum\limits_{i=0}^k  \left(\prod_{j=1}^{k-i} \lambda_{j}\right) [R_z z_{{\rm r},i}^{2}(\hat{\theta}) +  R_{\rmd} (\Phi_i\hat{\theta})^2] + \left(\prod_{j=1}^{k} \lambda_{j}\right) (\hat{\theta} - \theta_0)^\rmT R_{\theta} (\hat{\theta} - \theta_0),
    \label{costf}
\end{align}
where $R_z\in(0,\infty)$, $R_d\in(0,\infty)$, $\lambda_k \in (0, 1]$ is the forgetting factor, and the regularization weighting matrix $R_{\theta}\in\BBR^{l_{\theta} \times l_{\theta}}$ is positive definite.
Then, for all $k\ge 0$, the unique global minimizer 
\begin{equation}
\theta_{k+1} \triangleq \argmin_{\hat{\theta} \in \BBR^{l_\theta}} \SJ_k(\hat{\theta}) 
\end{equation}
is given recursively by the RLS update equations \mycite{islam2019recursive, lai2022exponential}
\vspace{-0.5em}
\begin{align}
P_{k+1}^{-1} &= \lambda_{k}P_k^{-1} + (1-\lambda_k)R_\infty + \widetilde{\Phi}_k^\rmT \widetilde{R} \widetilde{\Phi}_k, \label{covariance_update} \\
\theta_{k+1} &= \theta_{k} - P_{k+1} \widetilde{\Phi}^{\rmT}_{k} \widetilde{R} (\widetilde{z}_{k} + \widetilde{\Phi}_{k} \theta_{k}), \label{theta_update}
\end{align}
where $P_0 \isdef R_\theta^{-1}$, for all $k \ge 0$, $P_k \in \BBR^{l_\theta \times l_\theta}$ is the positive-definite covariance matrix, the positive-definite matrix $R_\infty \in \BBR^{l_\theta \times l_\theta}$ is the user-selected \textit{resetting matrix}, and where, for all $k \ge 0$,
\vspace{-0.5em}
\begin{gather*}
\widetilde{\Phi}_k \isdef \begin{bmatrix}
   \Phi_{\rmf, k}  \\
   \Phi_k   \\
\end{bmatrix}, \quad 
\widetilde{z}_k \isdef \begin{bmatrix}
   z_k-\hat{d}_{{\rm f},k}  \\
   0   \\
\end{bmatrix}, \quad
\widetilde{R} \isdef \begin{bmatrix}
   R_z & 0  \\
   0 & R_{\rmd}   \\
\end{bmatrix}.
\end{gather*}
Hence, \eqref{covariance_update} and \eqref{theta_update} recursively update the estimated coefficient vector \eqref{est_coeff_vec}.

The forgetting factor $\lambda_k \in(0,1]$ in \eqref{costf} and \eqref{covariance_update} enables the eigenvalues of $P_k$ to increase, 
which facilitates  adaptation of the input-estimation subsystem \eqref{estimate_law1}  %, even after extensive data collection ---not precise
\mycite{aastrom1977theory}.
In addition, the resetting matrix $R_\infty$ in \eqref{covariance_update} prevents the eigenvalues of $P_k$ from becoming excessively large under conditions of poor excitation \mycite{lai2022exponential}, a phenomenon known as covariance windup \mycite{malik1991some}.

Next, variable-rate forgetting based on the \textit{F}-test \mycite{mohseni2022recursive} is used to select the forgetting factor $\lambda_k \in (0,1]$.
For all $k \ge 0$, we define the \textit{residual error} at step $k$ by
\begin{align}
    \varepsilon_k \isdef \widetilde{z}_{k} + \widetilde{\Phi}_{k} \theta_{k} \in \mathbb{R}^2.
\end{align}
The residual error indicates how well the input-estimation subsystem \eqref{estimate_law1} predicts the input one step into the future.
Furthermore, for all $k \ge 0$, the sample mean of the residual errors over the previous $\tau \ge 1$ steps is defined by
\begin{align}
    \bar{\varepsilon}_{\tau,k} \isdef \frac{1}{\tau} \sum_{i=k - \tau + 1}^k {\varepsilon_i} \in \mathbb{R}^2,
\end{align}
\normalsize
and the sample variance of the residual errors over the previous $\tau$ steps is defined by
\begin{align}
    \Sigma_{\tau,k} \isdef \frac{1}{\tau}  \sum_{i=k - \tau + 1}^k (\varepsilon_i - \bar{\varepsilon}_{\tau,k})(\varepsilon_i - \bar{\varepsilon}_{\tau,k})^\mathrm{T} \in \mathbb{R}^{2 \times 2}.
\end{align}
\normalsize

The approach in \mycite{mohseni2022recursive} compares $\Sigma_{\tau_n,k}$ to $\Sigma_{\tau_d,k}$, where $\tau_n \ge 1$ is the short-term sample size, and $\tau_d > \tau_n$ is the long-term sample size.
If the short-term variance $\Sigma_{\tau_n,k}$ is found to be statistically more significant than the long-term variance $\Sigma_{\tau_d,k}$, according to the Lawley-Hotelling trace approximation \mycite{mckeon1974f}, then $\lambda_k < 1$ is chosen to be inversely proportional to its statistical significance.
Otherwise, $\lambda_k$ is set to 1.
In particular, for all $k \ge 0$, the forgetting factor is selected as

\begin{align}
    \lambda_k \isdef \frac{1}{1 + \eta g_k \mathbf{1}[g_k]}, \label{VRF_eq}
\end{align}
where $\eta \ge 0$ is a tuning parameter, $\mathbf{1} \colon \BBR \rightarrow \{0,1\}$ is the unit step function, and, for all $k \ge 0$,
\begin{align}
    g_k &\isdef \sqrt{  \frac{\tau_n}{\tau_d} \frac{\tr(\Sigma_{\tau_n,k} \Sigma_{\tau_d,k}^{-1})}{c} }
    - \sqrt{F_{2 \tau_n,b}^{-1} (1-\alpha)},
    \\
    a &\isdef \frac{(\tau_n + \tau_d - 3)(\tau_d -1)}{(\tau_d - 5)(\tau_d - 2)},
    \\
    b &\isdef 4 + \frac{2(\tau_n + 1)}{a-1}, \quad c \isdef \frac{2\tau_n(b-2)}{b(\tau_d - 3)},
\end{align}
and where $\alpha \in [0,1]$ is the significance level and $F_{2 \tau_n,b}^{-1} \colon [0,1] \rightarrow \BBR$ is the inverse cumulative distribution function of the \textit{F}-distribution with degrees of freedom $2 \tau_n$ and $b$. 
For further details, see \mycite{mohseni2022recursive} and \mycite{mckeon1974f}.

%  %

\subsection{State Estimation}

The forecast variable $x_{{\rm fc},k}$ updated by \eqref{kalman_fc_state} is used to obtain the estimate $x_{{\rm da},k}$ of $x_k$ given, for all $k \ge 0$, by the Kalman filter data-assimilation step
\begin{align}
x_{{\rm da},k} &= x_{{\rm fc},k} + K_{{\rm da},k} z_k, \label{kalman_da_state}
\end{align}
where the Kalman filter gain $K_{{\rm da},k} \in \mathbb R^{n}$, the data-assimilation error covariance $P_{{\rm da},k} \in \mathbb R^{n \times n},$
and the forecast error covariance $P_{\rmf\rmc,k+1} \in \mathbb R^{n \times n}$ are given by
\begin{align}
    K_{{\rm da},k} &= - P_{\rmf\rmc,k}C^{\rmT} ( C P_{\rmf\rmc,k} C^{\rmT} + V_{2,k}) ^{-1}, \label{kalman_gain} \\
    P_{{\rm da},k} &=  (I_{n}+K_{{\rm da},k}C) P_{\rmf\rmc,k},\label{Pda} \\
	P_{\rmf\rmc,k+1} &=  A P_{{\rm da},k}A^{\rmT} + V_{1,k}, \label{Pf}
\end{align}
where $V_{2,k} \in \mathbb R$ is the measurement noise covariance, $V_{1,k}$ is defined by 
\begin{align}
    V_{1,k}\isdef\,  B {\rm var}(d_k-\hat{d}_k)B^\rmT 
     + A {\rm cov}(x_k - x_{{\rm da},k},d_k-\hat{d}_k)B^\rmT 
     + B {\rm cov}(d_k-\hat{d}_k,x_k - x_{{\rm da},k})A^\rmT,
\end{align}
and $P_{\rmf\rmc,0} = 0.$ 
\subsection{Adaptive State Estimation} \label{sec:AdapInptStateEst}

This section summarizes the adaptive state estimation component of AISE. 
Assuming that, for all $k \ge 0$, $V_{1,k}$ and $V_{2,k}$ are unknown in (\ref{Pf}) and \eqref{kalman_gain},
the goal is to adapt ${V}_{{1,\rm adapt},k}$ and ${V}_{{2,\rm adapt},k}$ at each step $k$ to estimate $V_{1,k}$ and $V_{2,k}$, respectively.
To do this, we define, for all $k \ge 0$, the %computable 
performance metric $J_k \colon \BBR^{n\times n} \times \BBR \rightarrow \BBR$ by
\begin{align}
  {J}_{k}({V}_{1},{V}_{2}) \isdef |\widehat{S}_{ k}-{S}_{ k}|, \label{J_daptmetric}
\end{align}
where $\widehat{S}_{ k}$ is the sample variance of $z_k$ over $[0,k]$ defined by
\begin{align}
    \widehat{S}_{k} \isdef \cfrac{1}{k}\sum^{k}_{i=0}(z_i - \overline{z}_k)^2, \quad
    \overline{z}_k \isdef \cfrac{1}{k+1}\sum^{k}_{i=0}z_i,  \label{var_comp}
\end{align}
and ${S}_{k}$ is the variance of the residual $z_k$ determined by the Kalman filter, given by
\begin{align}
    % {S}_{k} \isdef  C P_{{\rm f},k} C^{\rm T} + V_{2,k}.  \label{var_inno}
    {S}_{k} \isdef  C (A P_{{\rm da},k-1}A^{\rmT} + V_{1}) C^{\rm T} + V_{2}.  \label{var_inno}
\end{align}
For all $k \ge 0$, we assume for simplicity that ${V}_{{1,\rm adapt},k}  \triangleq \eta_k I_n$, and we define the set $\SSS$ of minimizers 
$(\eta_k,{V}_{{2,\rm adapt},k})$ of $J_k$ by
\begin{align}
      \SSS\isdef \{ (\eta_k,{V}_{{2,\rm adapt},k}) \colon \eta \in [\eta_{\rmL},\eta_{\rmU}] \mbox{ and } {V}_{2} \ge 0 \mbox{ minimize }  J_k(\eta I_{n},V_{2})\}, \label{covmin}
\end{align} 
where
$0 \le \eta_{\rmL} \le \eta_{\rmU}.$
Next, defining ${J}_{\rmf,k} \colon \BBR \rightarrow \BBR $ by
\begin{align}
    {J}_{\rmf,k}(V_{1}) \isdef \widehat{S}_{k} - C (A P_{{\rm da},k-1}A^{\rmT} + V_{1})  C^{\rm T}, \label{J1_func}
\end{align}
and using \eqref{var_inno}, it follows that (\ref{J_daptmetric}) can be written as
\begin{align}
    {J}_k({V}_{1},{V}_{2}) = |{J}_{\rmf,k}(V_{1})-V_{2}|. \label{J_daptmetric_V2}
\end{align}
We then construct the set $\SJ_{\rmf,k}$ of positive values of ${J}_{\rmf,k}$ given by % by enumerating ${V}_{1,k-1} = \eta I_{n}$ as
\begin{align}
      \SJ_{\rmf,k} \isdef \{J_{\rmf,k}(\eta I_{n}) \colon J_{\rmf,k}(\eta I_{n}) > 0, \eta_{\rmL} \le\eta \le\eta_{\rmU}\} \subseteq \BBR. \label{J_f_positive}
\end{align}
Following result provides a technique for computing $\eta_k$ and ${V}_{{2,\rm adapt},k}$ defined in \eqref{covmin}.

\begin{prop}\label{prop: eta_k and V2,adapt minimizer}
    Let $k \ge 0$.  Then, the following statements hold:
    \begin{enumerate}
        \item  Assume that $\SJ_{\rmf,k}$ is nonempty, let $\beta \in [0,1]$, and define $\eta_k$ and $V_{2,k}$ by
    \begin{gather}
        \eta_k = \underset{\eta \in [\eta_L,\eta_U]}{\arg \min} \ |J_{\rmf,k}(\eta I_{n}) -  \widehat{J}_{\rmf,k}(\beta)|,\\
        {V}_{{2,\rm adapt},k} = J_{\rmf,k}(\eta_k I_n),
        \label{v_2_opt_1_NE}
    \end{gather}
    where
\begin{align}
       \widehat{J}_{\rmf,k}(\beta) \isdef \beta \min \SJ_{\rmf,k}+(1-\beta)\max \SJ_{\rmf,k}. \label{alpha1}
\end{align}
Then, $(\eta_k,{V}_{{2,\rm adapt},k})\in\SSS.$
\item      Assume that $\SJ_{\rmf,k}$ is empty, and define $\eta_k$ and $V_{2,k}$ by
\begin{gather}
        \eta_k = \underset{\eta \in [\eta_\rmL,\eta_\rmU]}{\arg \min} \ |J_{\rmf,k}(\eta I_{n})|,\\
        {V}_{{2,\rm adapt},k} = 0. \label{v_2_opt_1_E}
\end{gather}
Then, $(\eta_k,{V}_{{2,\rm adapt},k})\in\SSS.$
    \end{enumerate}
\end{prop}

\textit{Proof:} See Section 5.2 of \mycite{verma_shashank_2023_realtime_IJC}.

\begin{figure}[h!t]
  \begin{center}
{\includegraphics[width=0.75\linewidth]{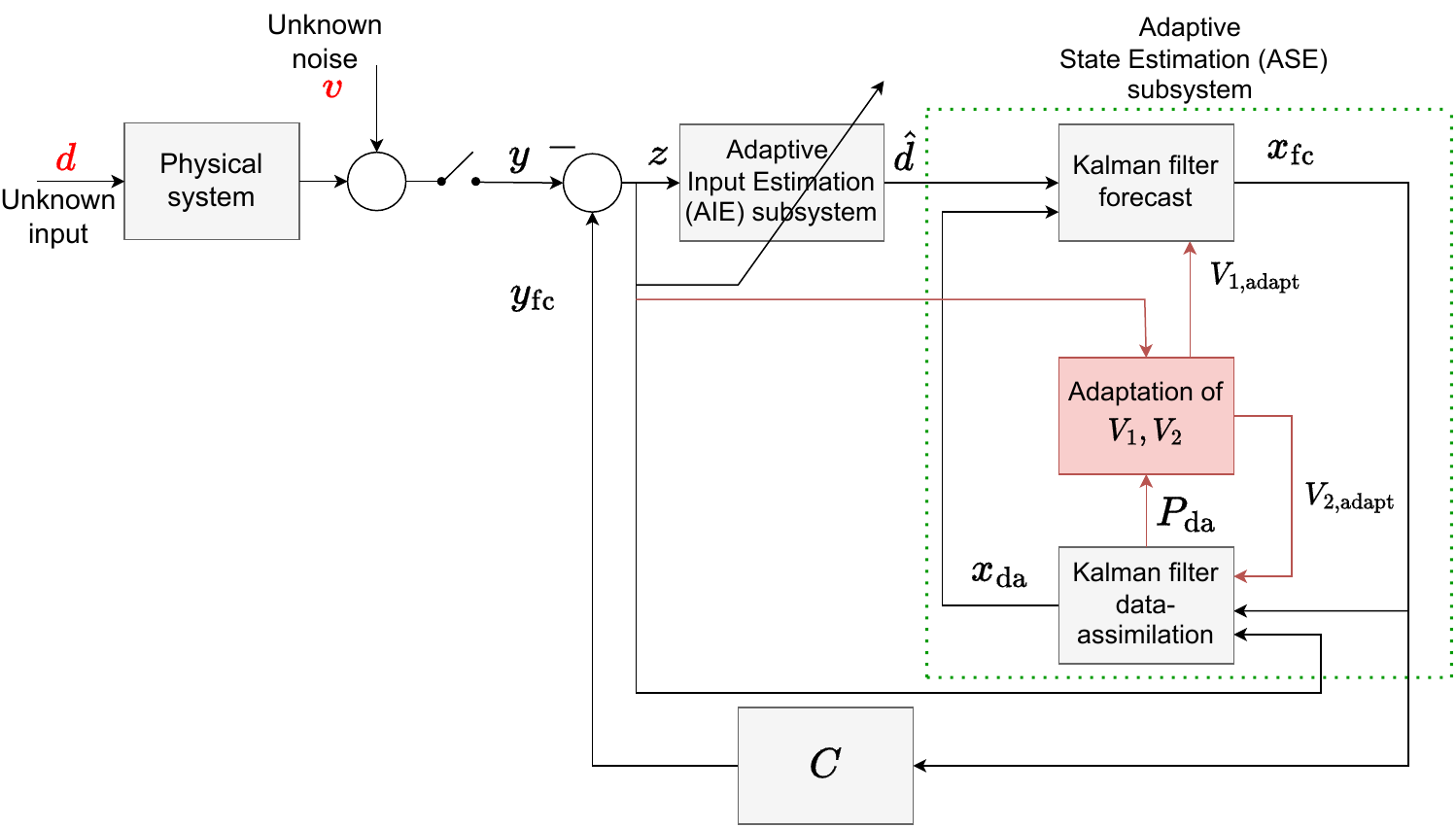}}
\end{center}
\caption{Block diagram of AISE.}
\label{fig:sec5_block_diag_AIE_ASE}
\end{figure}

A block diagram of AISE is shown in Figure \ref{fig:sec5_block_diag_AIE_ASE}. 

%%%%%%%%%%%%%%%%%%%%%%%%%%%%%%%%%%%%%%%%%%%%%%%%%%%%%%%%%%%
\section{Sensor-Fault Detection} \label{sec:sensor_fault_Detection}
KSFD computes each term in \eqref{acc_ddotE}, \eqref{single_transport_resolve_B_long}, and \eqref{double_transport_resolve_B_long} using sensor data and derivatives of sensor data. 
The equation residuals are then used to detect sensor faults.
We denote the LHS of \eqref{single_transport_resolve_B_long} by $ \begin{bmatrix}
    L_{\rms,{x},k} & L_{\rms,{y},k} & L_{\rms,{z},k}
\end{bmatrix}^\rmT $, the LHS of \eqref{double_transport_resolve_B_long} by $ \begin{bmatrix}
    L_{\rmd,{x},k} & L_{\rmd,{y},k} & L_{\rmd,{z},k}
\end{bmatrix}^\rmT $,
and the RHS of \eqref{acc_ddotE}, \eqref{single_transport_resolve_B_long}, and \eqref{double_transport_resolve_B_long} by $ \begin{bmatrix}
    R_{\rma,{x},k} & R_{\rma,{y},k} & R_{\rma,{z},k}
\end{bmatrix}^\rmT $, $ \begin{bmatrix}
    R_{\rms,{x},k} & R_{\rms,{y},k} & R_{\rms,{z},k}
\end{bmatrix}^\rmT $, and $ \begin{bmatrix}
    R_{\rmd,{x},k} & R_{\rmd,{y},k} & R_{\rmd,{z},k}
\end{bmatrix}^\rmT $ respectively.
Letting $\delta\ge1$ denote the window size, we define the error metrics for $ k > \delta$ by
\begin{equation}
    e_{\rms,{q},k} \isdef \sqrt{\frac{1}{\delta}\displaystyle\sum_{i=k-\delta}^{k}(L_{\rms,{q},i}-R_{\rms,{q},i})^2},  \label{rms_S}
\end{equation}
\begin{equation}
    e_{\rmd,{q},k} \isdef \sqrt{\frac{1}{\delta}\displaystyle\sum_{i=k-\delta}^{k}(L_{\rmd,{q},i}-R_{\rmd{,q},i})^2},  \label{rms_D}
\end{equation}
\begin{equation}
    e_{\rma,{q},k} \isdef \sqrt{\frac{1}{\delta}\displaystyle\sum_{i=k-\delta}^{k}(L_{\rmd,{q},i}-R_{\rma,{q},i})^2},  \label{rms_A}
\end{equation}
where $q$ represents ${x},$ $y$, or $z$. 
We consider the following types of sensor faults:  
\begin{itemize}
    \item {\it Bias}: An offset is added to the sensor measurement.
    \item {\it Drift}: A ramp is added to the sensor measurement. 
    \item {\it Sinusoid}: A sinusoid is added to the sensor measurement.
    \item {\it Severe noise}: A high level of white noise is added to the sensor measurement. 
\end{itemize}

%%%%%%%%%%%%%%%%%%%%%%%%%%%%%%%%%%%%%%%%%%%%%%%%%%%%%%%%%%%
\subsection{Sensor-Fault Detection for Ground Vehicles} \label{sec:sensor_fault_detection_ground_vehicle}
Here we apply KSFD to ground vehicles that operate on the horizontal plane.
We view the radar as a single sensor providing distance data along the body $x$ and $y$ directions.
Under this assumption, there are five relevant sensors, namely, the magnetometer, the radar, the $z$ rate gyro, and the $x$ and $y$ accelerometer.
The relevant error metrics are
${e_{\rms,{x},k}}, {e_{\rms,{y},k}}, {e_{\rmd,{x},k}},$ $ {e_{\rmd,{y},k}}, {e_{\rma,{x},k}},$ and ${e_{\rma,{y},k}}$.
Table \ref{Tab:Table_error_metric_sensor_ground} summarizes the sensors used to compute the error metrics \eqref{rms_S}--\eqref{rms_A} in the case of a ground vehicle which is restricted to horizontal plane.

\begin{center} 
\begin{tabular}{ |c|c| }%{ |c{2.5cm}||c{2.5cm}|c{2.5cm}|}
 \hline
 \textbf{Error Metric} & \textbf{Sensors} \\
 \hline
 $e_{\rms,{q},k}$ & Magnetometer, Radar, Rate gyro \\
 \hline
 $e_{\rmd,{q},k}$ & \makecell{Radar, Rate gyro, Accelerometers} \\
 \hline
 $e_{\rma,{q},k}$ & Magnetometer, Radar, Accelerometers\\
 \hline
\end{tabular}
\captionof{table}{\it Sensors used in KSFD to compute each error metric for the ground vehicle, where $q$ represents $x$ or $y$. } \label{Tab:Table_error_metric_sensor_ground}
\end{center}
 
Each error metric is computed in real time over a trailing data window as the vehicle operates. 
For each error metric, we specify a cutoff threshold, namely,
$c_{\rms,x}$, $c_{\rms,y}$, $c_{\rmd,x}$, $c_{\rmd,y}$, $c_{\rma,x}$, and $c_{\rma,y}$ for
${e_{\rms,{x},k}}$, ${e_{\rms,{y},k}}$, ${e_{\rmd,{x},k}}$, ${e_{\rmd,{y},k}}$, ${e_{\rma,{x},k}}$, and ${e_{\rma,{y},k}},$ respectively.
At each step $k,$ each error metric is either above cutoff (AC) or below cutoff (BC), and the goal is to use this information to detect faulty sensors.
To simplify this procedure, we assume that at most one sensor among the magnetometer, radar, $z$ rate gyro, and $x$ and $y$ accelerometers is faulty at any given time. The method for setting the cutoff thresholds is discussed in the next section.

Considering all possible values of the six error metrics, there are 64 potential scenarios.
Assuming that at most one sensor among the magnetometer, radar, $z$ rate gyro, $x$ accelerometer, and $y$ accelerometer is faulty, it follows that, within these 64 cases, only six are feasible.
For each of these six cases, Table  \ref{Tab:Sensor_Diagonastic_ground_vehicle} specifies the cutoff information that identifies the faulty sensor.

\begin{table*}[h!t]
\begin{center} 
\begin{tabular}{ |c|c|c|c|c|c|c| }%{ |c{2.5cm}||c{2.5cm}|c{2.5cm}|}
 \hline
 ${e_{\rms,{x},k}}$ & ${e_{\rms,{y},k}}$ & ${e_{\rmd,{x},k}}$ & ${e_{\rmd,{y},k}}$ & ${e_{\rma,{x},k}}$ & ${e_{\rma,{y},k}}$ & \textbf{Diagnostic} \\
 \hline
   BC & BC & BC & BC & BC & BC & All sensors are healthy\\
 \hline
  AC & AC & BC & BC & AC & AC & Magnetometer is faulty\\
 \hline
  AC & AC & AC & AC & AC & AC & Radar is faulty\\
 \hline
  AC & AC & AC & AC & BC & BC & $z$ Rate gyro is faulty\\
 \hline
  BC & BC & AC & BC & AC & BC & $x$ Accelerometer is faulty\\
 \hline
  BC & BC & BC & AC & BC & AC & $y$ Accelerometer is faulty\\
 \hline
 % - & - & - & \makecell{Rest of the 4 cases \\ are not possible}\\
 %  \hline
\end{tabular}
\captionof{table}{\it Identification of sensor faults in KSFD for ground vehicles confined to the horizontal plane.  At each step, all six error metrics are computed using data obtained during a trailing window.  For each metric, BC indicates that the error metric is below cutoff, and AC indicates that the error metric is above cutoff.   It is assumed that at most one sensor is faulty; under this assumption, the remaining 58 cases cannot occur. 
 }\label{Tab:Sensor_Diagonastic_ground_vehicle}
\end{center} 
\end{table*}

%%%%%%%%%%%%%%%%%%%%%%%%%%%%%%%%%%%%%%%%%%%%%%%%%%%%%%%%%%%
\subsection{Sensor-Fault Detection for Aerial Vehicles} \label{sec:sensor_fault_detection_Ariel_vehicle}

Here we apply KSFD to aerial vehicles.
We consider either radar as a single sensor providing distance data along the body $x$, $y$, and $z$ directions. 
We assume that an IMU is responsible for determining the attitude of the vehicle at all times.
Under this assumption, there are eight relevant sensors, namely, the IMU, the radar, the $x$, $y$, and $z$ rate gyros, and the $x$, $y$, and $z$ accelerometers.
The relevant error metrics are
${e_{\rms,{x},k}}, {e_{\rms,{y},k}}, {e_{\rms,{z},k}}, {e_{\rmd,{x},k}},$ $ {e_{\rmd,{y},k}}, {e_{\rmd,{z},k}}, {e_{\rma,{x},k}}, {e_{\rma,{y},k}},$ and ${e_{\rma,{z},k}}$.
Table \ref{Tab:Table_error_metric_sensor_aerial} summarizes the sensors used to compute the error metrics \eqref{rms_S}--\eqref{rms_A} in the case of a aerial vehicle.

\begin{center} 
\begin{tabular}{ |c|c| }%{ |c{2.5cm}||c{2.5cm}|c{2.5cm}|}
 \hline
 \textbf{Error Metric} & \textbf{Sensors} \\
 \hline
 $e_{\rms,{q},k}$ & IMU, Radar, Rate gyro \\
 \hline
 $e_{\rmd,{q},k}$ & \makecell{Radar, Rate gyro, Accelerometers} \\
 \hline
 $e_{\rma,{q},k}$ & IMU, Radar, Accelerometers\\
 \hline
\end{tabular}
\captionof{table}{\it Sensors used in KSFD to compute each error metric for the aerial vehicle, where $q$ represents $x,$ $y,$ or $z$. } \label{Tab:Table_error_metric_sensor_aerial}
\end{center}

Similar to Section \ref{sec:sensor_fault_detection_ground_vehicle}, each error metric in KSFD is computed in real time over a trailing data window as the vehicle operates. 
For each error metric, we specify a cutoff threshold, namely,
$c_{\rms,x}$, $c_{\rms,y}$, $c_{\rms,z}$, $c_{\rmd,x}$, $c_{\rmd,y}$, $c_{\rmd,z}$, $c_{\rma,x}$, $c_{\rma,y}$, and $c_{\rma,z}$ for
${e_{\rms,{x},k}}$, ${e_{\rms,{y},k}}$, ${e_{\rms,{z},k}}$, ${e_{\rmd,{x},k}}$, ${e_{\rmd,{y},k}}$, ${e_{\rmd,{z},k}}$, ${e_{\rma,{x},k}}$, ${e_{\rma,{y},k}}$,and ${e_{\rma,{y},z}},$ respectively.
At each step $k,$ each error metric is either above cutoff (AC) or below cutoff (BC), and the goal is to use this information to detect faulty sensors.
To simplify this procedure, we assume that at most one sensor among the IMU, radar, $x$, $y$, and $z$ rate gyros, and the $x$, $y$, and $z$ accelerometer is faulty at any given time. 
The method for setting the cutoff thresholds is discussed in the next section.

Considering all possible values of the nine error metrics, there are 512 potential scenarios.
Assuming that at most one sensor among the IMU, radar, $x, y$, and $z$ rate gyros, $x,y,$ and $z$ accelerometers is faulty, it follows that, within these 512 cases, only nine are feasible.
For each of these nine cases, Table  \ref{Tab:Sensor_Diagonastic_aerial_vehicles} specifies the cutoff information that identifies the faulty sensor.

\begin{table*}[h!t]
\begin{center} 
\vspace{+0.04in}
\begin{tabular}{ |c|c|c|c|c|c|c|c|c|c| }%{ |c{2.5cm}||c{2.5cm}|c{2.5cm}|}
 \hline
 ${e_{\rms,{x},k}}$ & ${e_{\rms,{y},k}}$ & ${e_{\rms,{z},k}}$ & ${e_{\rmd,{x},k}}$ & ${e_{\rmd,{y},k}}$ & ${e_{\rmd,{z},k}}$ &${e_{\rma,{x},k}}$ & ${e_{\rma,{y},k}}$ & ${e_{\rma,{z},k}}$ & \textbf{Diagnostic} \\
 \hline
  BC & BC & BC & BC & BC & BC & BC & BC & BC & All sensors are healthy\\
 \hline
  AC & AC & AC & BC & BC & BC & AC & AC & AC & IMU is faulty\\
 \hline
  AC & AC & AC & AC & AC & AC & AC & AC & AC & Radar is faulty\\
 \hline
  AC & AC & BC & AC & AC & AC & BC & BC & BC & $z$ Rate gyro is faulty\\
 \hline
  AC & BC & AC & AC & AC & AC & BC & BC & BC & $y$ Rate gyro is faulty\\
 \hline
  BC & AC & AC & AC & AC & AC & BC & BC & BC & $x$ Rate gyro is faulty\\
 \hline
  BC & BC & BC & BC & BC & AC & BC & BC & AC & $z$ Accelerometer is faulty\\
 \hline
  BC & BC & BC & BC & AC & BC & BC & AC & BC & $y$ Accelerometer is faulty\\
 \hline
  BC & BC & BC & AC & BC & BC & AC & BC & BC & $x$ Accelerometer is faulty\\
 \hline
 % - & - & - & \makecell{Rest of the 4 cases \\ are not possible}\\
 %  \hline
\end{tabular}
\captionof{table}{\it Identification of sensor faults in KSFD for aerial vehicles.  At each step, all nine error metrics are computed using data obtained during a trailing window.  For each error metric, BC indicates that the error metric is below the cutoff, and AC indicates that the error metric is above the cutoff.   It is assumed that at most one sensor is faulty; under this assumption, the remaining 503 cases cannot occur. 
 }\label{Tab:Sensor_Diagonastic_aerial_vehicles}
\end{center} 
\vspace{-.3in}
\end{table*}

%%%%%%%%%%%%%%%%%%%%%%%%%%%%%%%%%%%%%%%%%%%%%%%%%%%%%%%%%%%
\section{Simulated Application to Ground Vehicle}  \label{sec:num_example_simu}

\begin{figure}[h!t]
              \begin{center}
            {\includegraphics[width=0.5\linewidth]{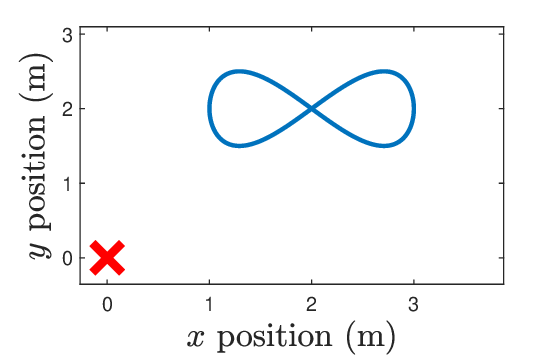}}
            \end{center}
            \vspace{-.15in}
            \caption{Position trajectory of the vehicle. The red $\times$ has zero inertial acceleration and serves as the radar target. The blue curve represents the simulated trajectory.} 
            \label{fig:sim_eight_curve_trajectory}
            \vspace{-.15in}
          \end{figure}

This section presents a numerical example to illustrate KSFD on the simulated data. The example addresses sensor fault specifically related to $x$-axis accelerometer drift. The analysis is confined to cases where motion is restricted to the horizontal (2D) plane.
To simulate measurements from the magnetometer, radar, $z$ rate gyro, and $x$ and $y$ accelerometers on a vehicle limited to the horizontal plane, we generate a figure-8 trajectory, as shown in Figure \ref{fig:sim_eight_curve_trajectory}.

The discrete-time equations governing the trajectory are defined by $x(k) = 2 + \sin(2 k T_\rms)$ and $y(k) = 2 + \sin(2 k T_\rms) \cos(2 k T_\rms)$, with the sampling interval $T_\rms = 0.01$ s/step for $k \in [1, 6000]$.
White Gaussian noise is introduced to simulate sensor noise, with standard deviations $\sigma_{r_x} = \sigma_{r_{y}} = 0.001$ m, $\sigma_{\omega_{z}} = 0.0001$ rad/s, and $\sigma_{A_{x}} = \sigma_{A_{y}} = 0.01g$, where $g = 9.8$ m/s$^2$.
In Example \ref{eg_sim_accelerometer_failure}, the transient phase for AISE is less than $1000$ steps. Therefore, we set $\delta = 1000$ steps.
The cutoff thresholds for each error metric are established at twice the rms error due to the sensor noise floor at step $k = 2\delta$. This setting ensures that the computations of the cutoff threshold at step $k = 2\delta$ are not influenced by the transient phase and by the fault starting from step $k = 3\delta$.

\begin{exam} \label{eg_sim_accelerometer_failure}

      {\it KSFD for Accelerometer Failure.}

      {\rm In this scenario, the $x$-axis accelerometer experiences a failure due to a drift with slope of $0.05$ $g$/$s$ starting at 30 s, where $g = 9.8 \rmm^2/s$. For single differentiation using AISE, we set $n_\rme = 25$, $n_\rmf = 50$, $R_z = 1$, $R_d = 10^{-6.7}$, $R_\theta = 10^{-8}I_{25}$, $\eta = 0.2, \tau_n = 5, \tau_d = 25, \alpha = 0.2$, and $R_{\infty} = 10^{-4}.$ The parameters $V_{1,k}$ and $V_{2,k}$ are adapted, with $\eta_{\rmL} = 10^{-6}$, $\eta_{\rmU} = 10^2$, and $\beta = 0.5$ as described in Section \ref{sec:AdapInptStateEst}. For double differentiation using AISE, we set $n_\rme = 20$, $n_\rmf = 18$, $R_z = 1$, $R_d = 10^{-5}$, $R_\theta = 10^{-8}I_{41}$, $\eta = 0.2, \tau_n = 5, \tau_d = 25, \alpha = 0.2$, and $R_{\infty} = 10^{-7}.$ Similarly, $V_{1,k}$ and $V_{2,k}$ are adapted, with $\eta_{\rmL} = 10^{-6}$,  $\eta_{\rmU} = 10^{-2}$, and $\beta = 0.5$ in Section \ref{sec:AdapInptStateEst}.

      Figure \ref{fig:sim_acc_DT_accelerometer_drift} compares $R_{\rmd,{x},k}$ and $R_{\rmd,{y},k}$ with their respective $L_{\rmd,{x},k}$ and $L_{\rmd,{y},k}$. In Figure \ref{fig:sim_error_DT_accelerometer_drift}, the drift begins at 30 s and leads to a gradual increase in $e_{\rmd,{x},k}$ above the cutoff $c_{\rmd,x}$, which indicates that one of the sensors used to compute $e_{\rmd,{x},k}$ is faulty.
      Figure \ref{fig:sim_acc_ST_accelerometer_drift} compares $R_{\rms,{x},k}$ and $R_{\rms,{y},k}$ with $L_{\rms,{x},k}$ and $L_{\rms,{y},k}$.
      Figure \ref{fig:sim_error_ST_accelerometer_drift} shows that $e_{\rms,{x},k}$ and $e_{\rms,{y},k}$ are below the cutoff threshold.
      Similarly, Figure \ref{fig:sim_acc_A_accelerometer_drift} and Figure \ref{fig:sim_error_A_accelerometer_drift} indicates that one of the sensors used to compute $e_{\rma,{x},k}$ is faulty.
      As observed, $e_{\rmd,{x},k}$ and $e_{\rma,{x},k}$ are above the cutoff thresholds, and thus Table \ref{Tab:Sensor_Diagonastic_ground_vehicle} implies that the $x$-axis accelerometer is faulty.

          \begin{figure}[h!t]
              \begin{center}
            {\includegraphics[width=0.7\linewidth]{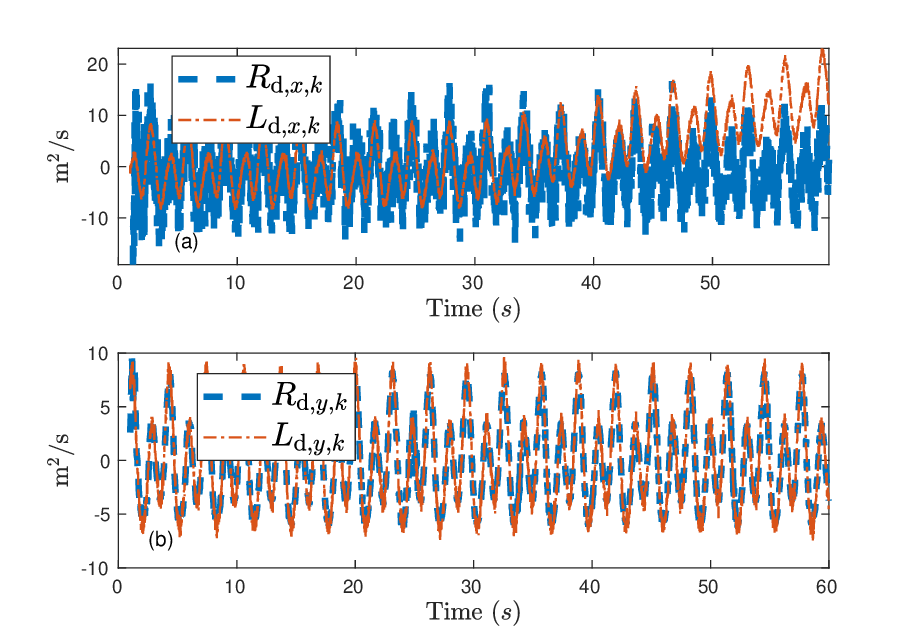}}
            \end{center}
            \caption{ {\it  Example \ref{eg_sim_accelerometer_failure}:}  Accelerometer with  drift. (a) Beginning at 30 s, $L_{\rmd,{x},k}$ exhibits drift. (b) $R_{\rmd,{y},k}$ follows $L_{\rmd,{y},k}$.} 
            \label{fig:sim_acc_DT_accelerometer_drift}
          \end{figure} 
          \begin{figure}[h!t]
              \begin{center}
            {\includegraphics[width=0.7\linewidth]{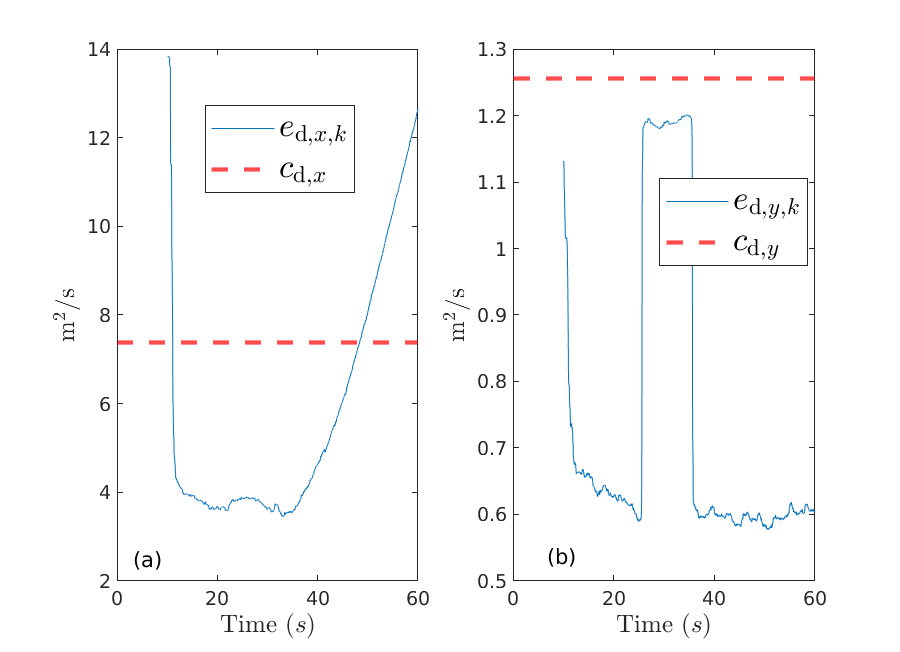}}
            \end{center}
            \caption{{\it  Example \ref{eg_sim_accelerometer_failure}:}  Accelerometer with  drift.  $e_{\rmd,{x},k}$ in (a) gradually increases at 30 s when the drift begins and exceeds the cutoff threshold $c_{\rmd,{x}}$, which indicates a fault in the sensors used to compute $e_{\rmd,{x},k}$. (b) $e_{\rmd,{y},k}$ remains below cutoff. } 
            \label{fig:sim_error_DT_accelerometer_drift}
          \end{figure} 
          \begin{figure}[h!t]
              \begin{center}
            {\includegraphics[width=0.7\linewidth]{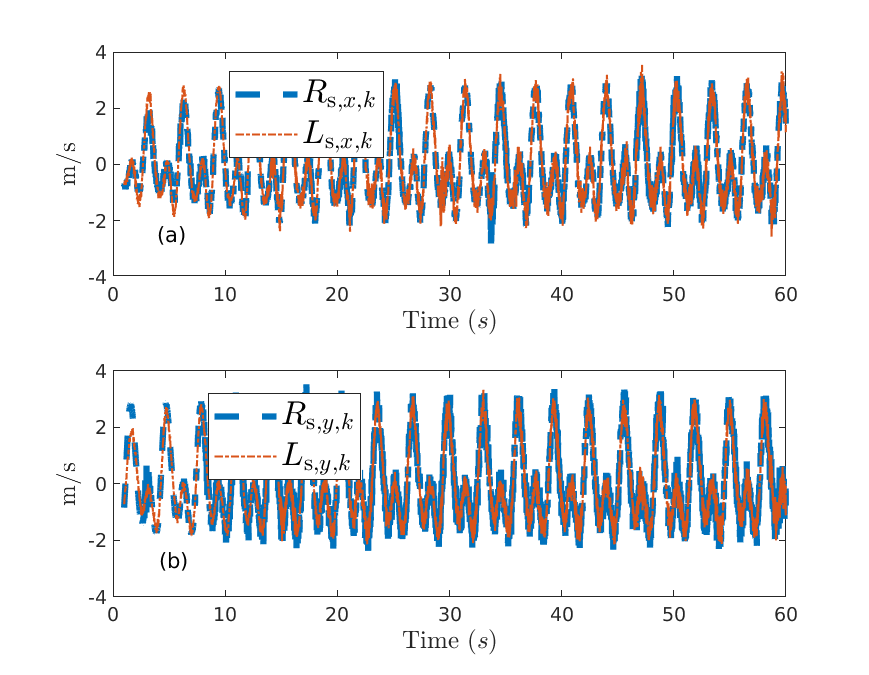}}
            \end{center}
            \caption{{\it  Example \ref{eg_sim_accelerometer_failure}:}  Accelerometer with drift. (a)  $R_{\rms,{x},k}$ follows $L_{\rms,{x},k}$. (b) $R_{\rms,{y},k}$ follows $L_{\rms,{y},k}$, which indicates no drift.} 
            \label{fig:sim_acc_ST_accelerometer_drift}
          \end{figure}
          \begin{figure}[h!t]
              \begin{center}  
              {\includegraphics[width=0.7\linewidth]{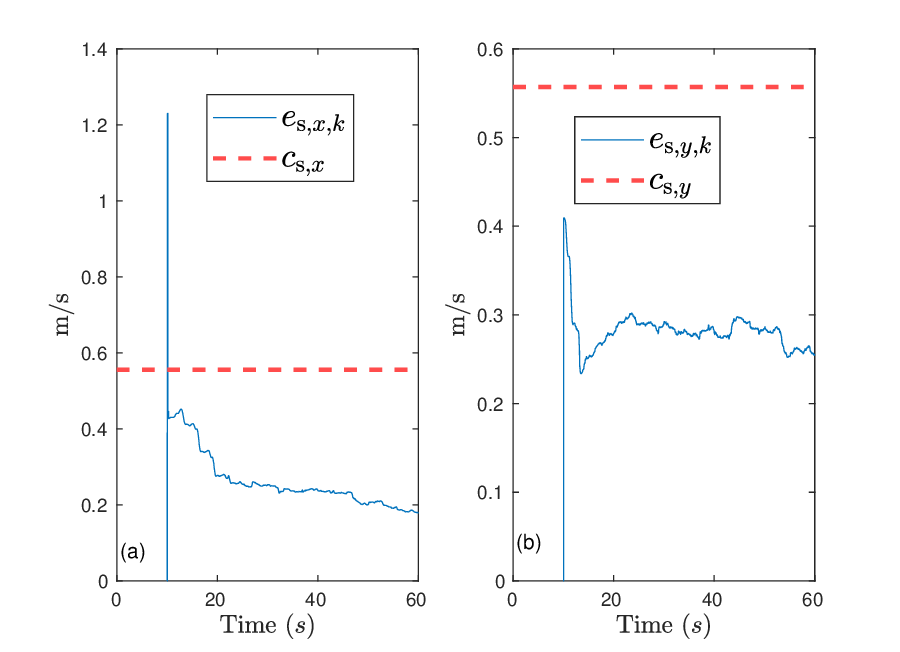}}
            \end{center}
            \caption{{\it  Example \ref{eg_sim_accelerometer_failure}:}  Accelerometer with drift. In (a) and (b), both $e_{\rms,{x},k}$ and $e_{\rms,{y},k}$ remain below the cutoff thresholds $c_{\rms,{x}}$ and $c_{\rms,{y}}$, respectively,  which indicates no sensor fault.} 
            \label{fig:sim_error_ST_accelerometer_drift}
          \end{figure}
          \begin{figure}[h!t]
              \begin{center}
            {\includegraphics[width=0.7\linewidth]{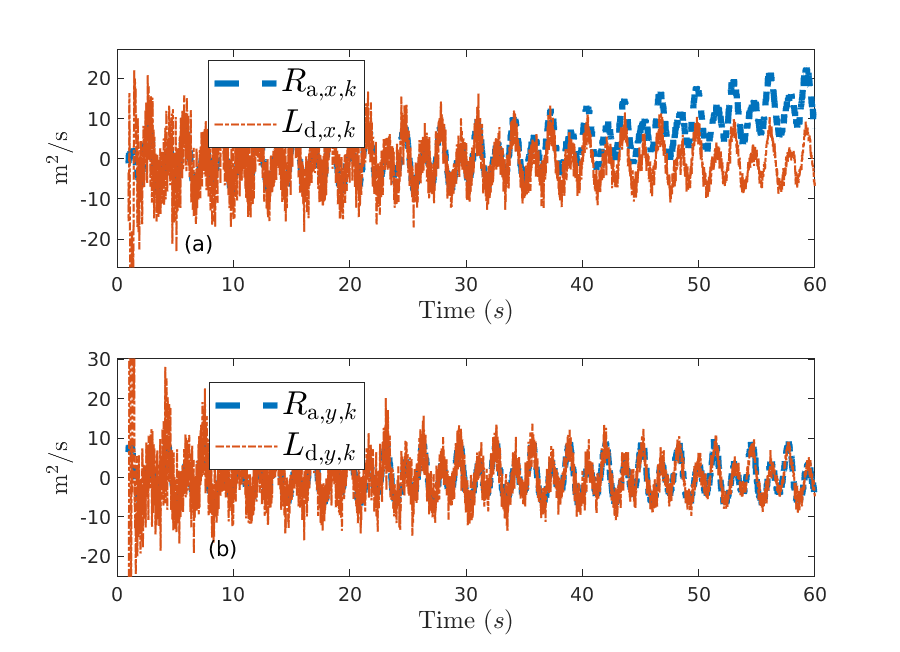}}
            \end{center}
            \vspace{-.1in}
            \caption{{\it  Example \ref{eg_sim_accelerometer_failure}:} Accelerometer with drift.  Beginning at 30 s, $R_{\rma,{x},k}$ in (a) exhibits drift. (b) $R_{\rma,{y},k}$ follows $L_{\rmd,{y},k}$.} 
            \label{fig:sim_acc_A_accelerometer_drift}
          \end{figure}
          \begin{figure}[h!t]
              \begin{center}
            {\includegraphics[width=0.7\linewidth]{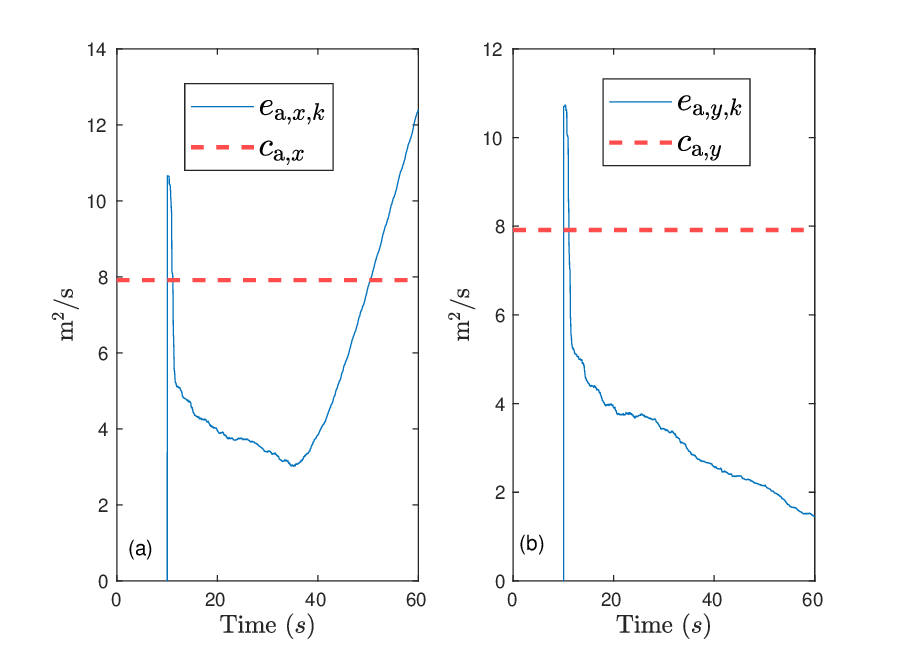}}
            \end{center}
            \caption{{\it  Example \ref{eg_sim_accelerometer_failure}:}  Accelerometer with drift.  $e_{\rma,{x},k}$ in (a) gradually increases at 30 s when the drift begins and exceeds the cutoff threshold $c_{\rma,{x}}$, which indicates a fault in the sensors used to compute $e_{\rma,{x},k}$. (b) $e_{\rma,{y},k}$ remains below cutoff.} 
            \label{fig:sim_error_A_accelerometer_drift}
          \end{figure}
    }      
    \end{exam}

%%%%%%%%%%%%%%%%%%%%%%%%%%%%%%%%%%%%%%%%%%%%%%%%%%%%%%%%%%%
\section{Experimental Application to Ground Vehicle}  \label{sec:num_example_exp_2d}
This section provides numerical examples of experimental data to illustrate KSFD in real-world scenarios. This section details the experimental results for a ground vehicle confined to the horizontal plane.

\begin{figure}
     \centering
     \begin{subfigure}[a]{0.6\textwidth}
         \centering
         \includegraphics[width=\textwidth]{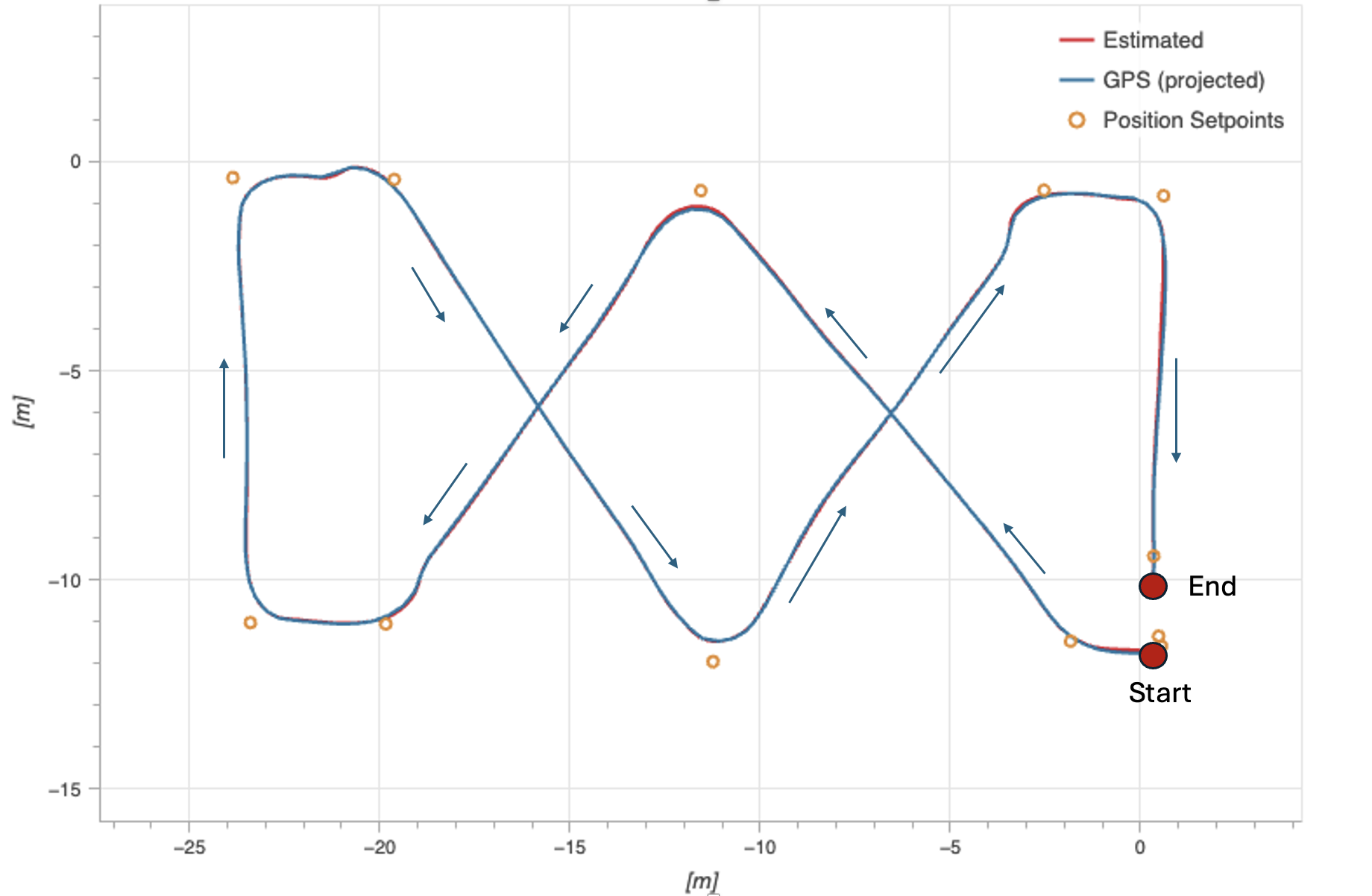}
         \caption{} 
            \label{fig:rover_trajectory}
     \end{subfigure}
     \hfill
     \begin{subfigure}[b]{0.4\textwidth}
         \centering
        \includegraphics[width=\textwidth]{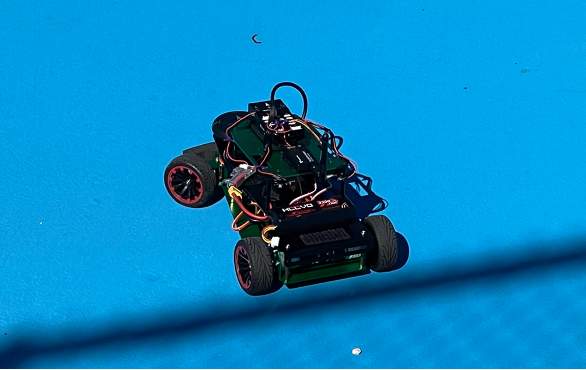}
         \caption{} 
            \label{fig:rover}
     \end{subfigure}
     % \hfill
     \caption{(a) Position trajectory of the ground vehicle. (b) Ackerman car chassis ground vehicle used in the experiment.}
        % \label{fig:three graphs}
\end{figure}

For the experimental setup, an Ackerman car chassis was constructed for the ground vehicle, equipped with a Pixhawk 6C autopilot. The ground rover traced a predefined zig-zag trajectory, set via the ground control application on a tennis court, illustrating the restriction to horizontal (2D) motion as shown in Figure \ref{fig:rover_trajectory}. The sensor data shown in Table \ref{Tab:SensorData_ground} were gathered from the Pixhawk 6C logs. Radar sensor faults attributed to high noise as well as $x$-axis accelerometer faults due to a sinusoid  are analyzed in examples \ref{eg_exp_radar_noise} and \ref{eg_exp_x_rate_sinu}, respectively.
For the experimental setup, an Ackerman car chassis was constructed for the ground vehicle, equipped with a Pixhawk 6C autopilot. The ground rover traced a predefined zig-zag trajectory, set via the ground control application on a tennis court, illustrating the restriction to horizontal (2D) motion as shown in Figure \ref{fig:rover_trajectory}. The sensor data shown in Table \ref{Tab:SensorData_ground} were gathered from the Pixhawk 6C logs. Radar sensor faults attributed to high noise and $x$-axis accelerometer faults due to a sinusoid are analyzed in examples \ref{eg_exp_radar_noise} and \ref{eg_exp_x_rate_sinu}, respectively.

In examples \ref{eg_exp_radar_noise} and \ref{eg_exp_x_rate_sinu}, the transient phase for AISE is concluded within $1000$ steps. Consequently, $\delta$ is set at $1000$ steps.
Cutoff thresholds are set at twice the error magnitude at step $k = 1.5\delta$. This parameter choice ensures that the data employed to compute the error metric at step $k = 1.5\delta$ are unaffected by the transient phase and by the fault starting from step $k = 2\delta$.

\begin{exam} \label{eg_exp_radar_noise}

      {\it KSFD for Radar failure.}

    {\rm In this scenario, the radar experiences a failure due to high white noise with a standard deviation of $1.0$ m, beginning at 20 s. The parameter settings are the same as those used in Example \ref{eg_sim_accelerometer_failure}.

    Figure \ref{fig:exp_acc_DT_radar_noise} compares $R_{\rmd,{x},k}$ and $R_{\rmd,{y},k}$ with their respective $L_{\rmd,{x},k}$ and $L_{\rmd,{y},k}$. Figure \ref{fig:exp_error_DT_radar_noise} demonstrates that the onset of high noise at 20 s causes an abrupt increase in $e_{\rmd,{x},k}$ and $e_{\rmd,{y},k}$, exceeding the cutoff thresholds $c_{\rmd,x}$ and $c_{\rmd,y}$, respectively, which indicates a fault in one of the sensors used to compute $e_{\rmd,{x},k}$ and $e_{\rmd,{y},k}$.
    Figure \ref{fig:exp_acc_ST_radar_noise} displays comparisons of $R_{\rms,{x},k}$ and $R_{\rms,{y},k}$ with $L_{\rms,{x},k}$ and $L_{\rms,{y},k}$. Likewise, Figure \ref{fig:exp_error_ST_radar_noise} illustrates that the high noise starting at 20 s results in significant increases in $e_{\rms,{x},k}$ and $e_{\rms,{y},k}$ above the cutoffs $c_{\rms,x}$ and $c_{\rms,y}$, respectively, signaling faults in the associated sensors.
    Figures \ref{fig:exp_acc_A_radar_noise} and \ref{fig:exp_error_A_radar_noise} further indicate sensor faults, as $e_{\rma,{x},k}$ and $e_{\rma,{y},k}$ also exceed their respective cutoff thresholds.
    Given that all error metrics exceed their cutoff thresholds, Table \ref{Tab:Sensor_Diagonastic_ground_vehicle} implies that the radar sensor is faulty.

          \begin{figure}[h!t]
              \begin{center}
            {\includegraphics[width=0.7\linewidth]{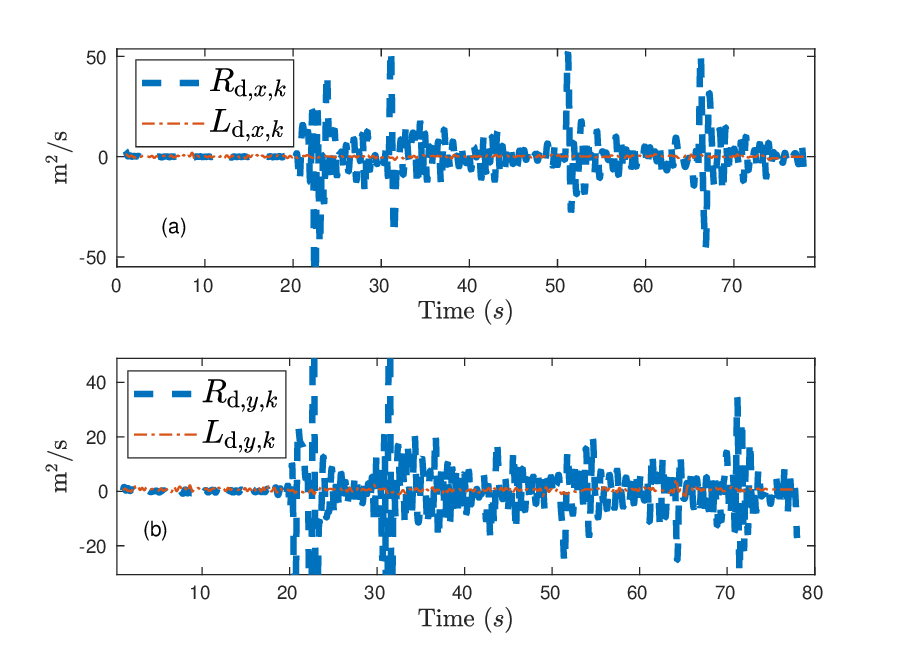}}
            \end{center}
            \caption{ {\it Example \ref{eg_exp_radar_noise}:} Radar data with high noise. Starting at 20 s, $R_{\rmd,{x},k}$ in (a) and $R_{\rmd,{y},k}$ in (b) exhibit high noise levels, respectively.}
            \label{fig:exp_acc_DT_radar_noise}
          \end{figure} 
          \begin{figure}[h!t]
              \begin{center}
            {\includegraphics[width=0.7\linewidth]{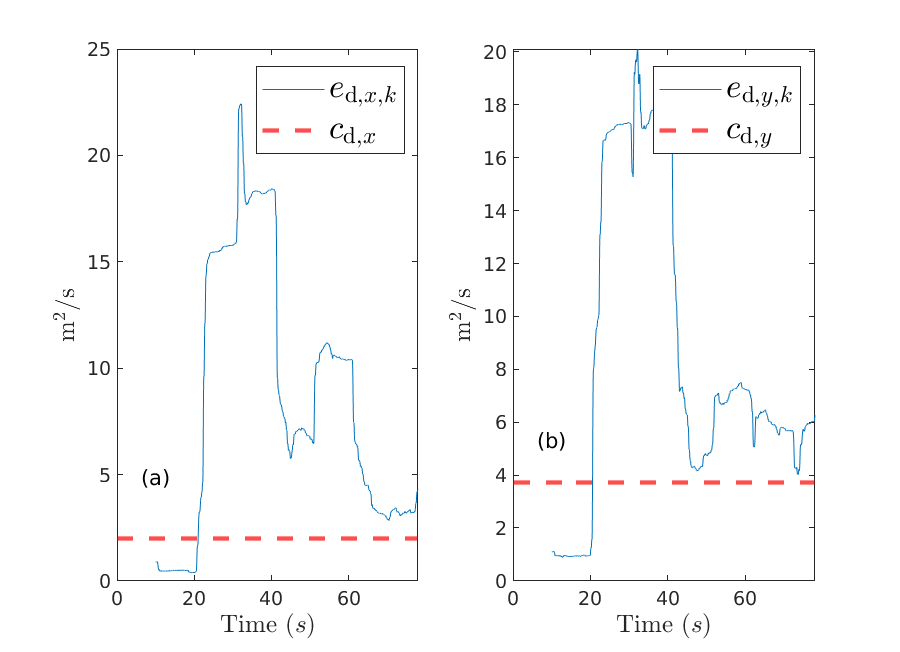}}
            \end{center}
            \caption{{\it  Example \ref{eg_exp_radar_noise}:}  Radar data with high noise.  $e_{\rmd,{x},k}$ in (a) and $e_{\rmd,{y},k}$ in (b) abruptly increase above the cutoff thresholds $c_{\rmd, x}$ and $c_{\rmd, y}$, respectively, at 20 s when the high noise begins, which indicates a fault in the sensors used to compute $e_{\rmd,{x},k}$ and $e_{\rmd,{y},k}$.} 
            \label{fig:exp_error_DT_radar_noise}
          \end{figure} 
          \begin{figure}[h!t]
              \begin{center}
            {\includegraphics[width=0.7\linewidth]{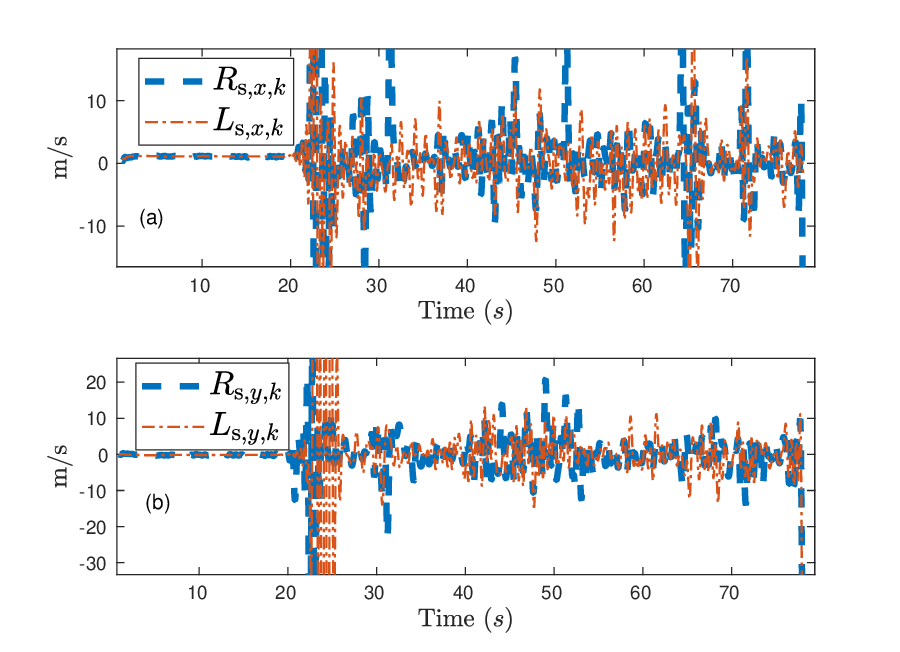}}
            \end{center}
            \caption{{\it  Example \ref{eg_exp_radar_noise}:}  Radar data with high noise. Starting at 20 s, $R_{\rms,{x},k}, L_{\rms,{x},k}$ in (a) and $R_{\rms,{y},k}, L_{\rms,{y},k}$ in (b) exhibit high noise levels, respectively.} 
            \label{fig:exp_acc_ST_radar_noise}
          \end{figure}
          \begin{figure}[h!t]
              \begin{center}  
              {\includegraphics[width=0.7\linewidth]{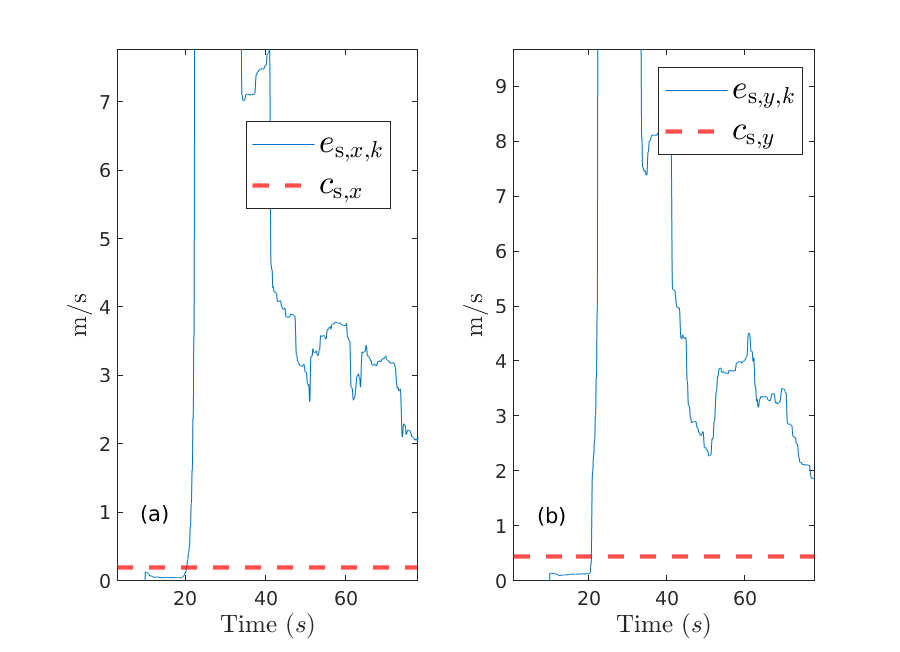}}
            \end{center}
            \caption{{\it  Example \ref{eg_exp_radar_noise}:}  Radar data with high noise. $e_{\rms,{x},k}$ in (a) and $e_{\rms,{y},k}$ in (b) abruptly increase above the cutoff thresholds $c_{\rms, x}$ and $c_{\rms, y}$, respectively, at 20 s when the high noise begins, which indicates a fault in the sensors used to compute $e_{\rms,{x},k}$ and $e_{\rms,{y},k}$.} 
            \label{fig:exp_error_ST_radar_noise}
          \end{figure}
          \begin{figure}[h!t]
              \begin{center}
            {\includegraphics[width=0.7\linewidth]{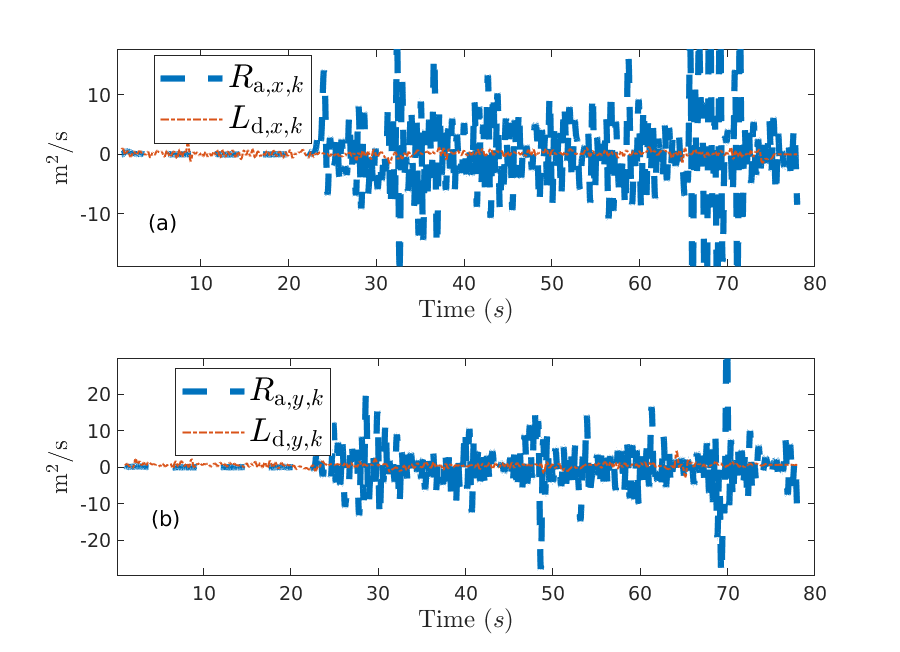}}
            \end{center}
            \caption{{\it  Example \ref{eg_exp_radar_noise}:} Radar with high noise.  Starting at 20 s, $R_{\rma,{x},k}$ in (a) and $R_{\rma,{y},k}$ in (b) exhibit high noise levels, respectively.} 
            \label{fig:exp_acc_A_radar_noise}
          \end{figure}
          \begin{figure}[h!t]
              \begin{center}
            {\includegraphics[width=0.7\linewidth]{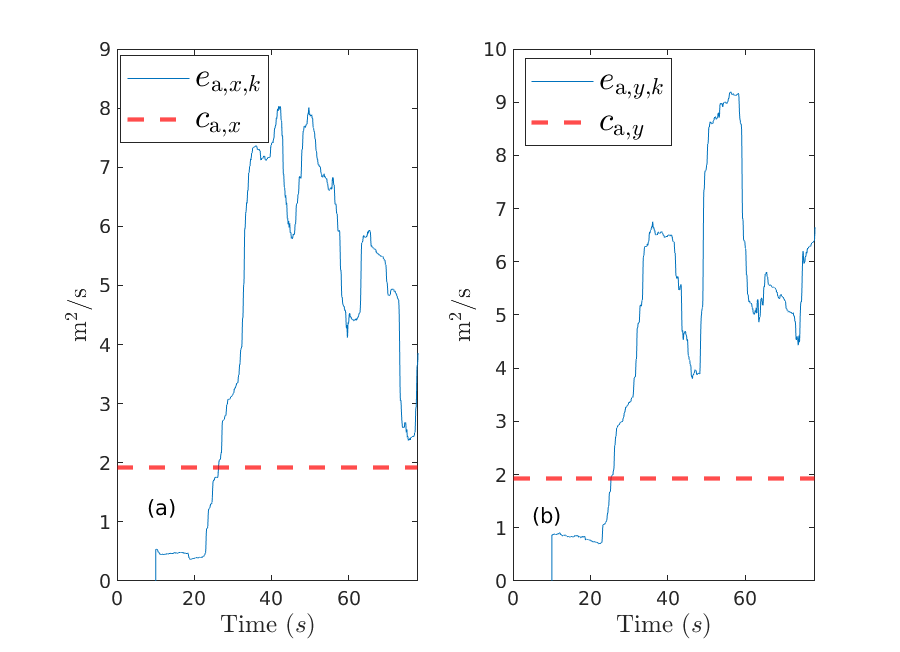}}
            \end{center}
            \caption{{\it  Example \ref{eg_exp_radar_noise}:}  Radar with high noise.  $e_{\rma,{x},k}$ in (a) and $e_{\rma,{y},k}$ in (b) abruptly increase above the cutoff thresholds $c_{\rma, x}$ and $c_{\rma, y}$, respectively, at 20 s when the high noise begins, which indicates a fault in the sensors used to compute $e_{\rma,{x},k}$ and $e_{\rma,{y},k}$.} 
            \label{fig:exp_error_A_radar_noise}
          \end{figure}
    }      
    \end{exam}

\begin{exam} \label{eg_exp_x_rate_sinu}

      {\it KSFD for rate-gyro failure.}

      {\rm In this scenario, the $z$-axis rate gyro encounters a failure due to the sinusoidal  sensor noise   $0.5\sin{(0.1kT_\rms)}$ starting at 30 s. The parameters for both single and double differentiation using AISE are the same as those used in Example \ref{eg_sim_accelerometer_failure}, except for $\alpha = 0.05$ for both single and double differentiation.

        Figure \ref{fig:exp_acc_DT_rate_sinu} compares $R_{\rmd,{x},k}$ and $R_{\rmd,{y},k}$ with their respective $L_{\rmd,{x},k}$ and $L_{\rmd,{y},k}$. Figure \ref{fig:exp_error_DT_rate_sinu} illustrates that the sinusoid fault, commencing at 30 s, results in a sharp increase in $e_{\rmd,{x},k}$ and  $e_{\rmd,{y},k}$ above the cutoff $c_{\rmd,x}$ and $c_{\rmd,y}$, suggesting a fault in one of the sensors contributing to $e_{\rmd,{x},k}$ and $e_{\rmd,{y},k}$.
        Figure \ref{fig:exp_acc_ST_rate_sinu} displays a comparison between $R_{\rms,{x},k}$ and $R_{\rms,{y},k}$, and $L_{\rms,{x},k}$ and $L_{\rms,{y},k}$. Similarly, Figure \ref{fig:exp_error_ST_rate_sinu} shows that $e_{\rms,{x},k}$ and $e_{\rms,{y},k}$ go above cutoff thresholds when the sensor fault begins at $30$ s.
        Figures \ref{fig:exp_acc_A_rate_sinu} and \ref{fig:exp_error_A_rate_sinu} indicates no sensor fault..
        Given that $e_{\rmd,{x},k}$, $e_{\rmd,{y},k}$, $e_{\rms,{x},k}$, and $e_{\rms,{y},k}$ exceed their respective cutoff thresholds, Table \ref{Tab:Sensor_Diagonastic_ground_vehicle} implies that the $z$-axis rate gyro is faulty.

          \begin{figure}[h!t]
              \begin{center}
            {\includegraphics[width=0.7\linewidth]{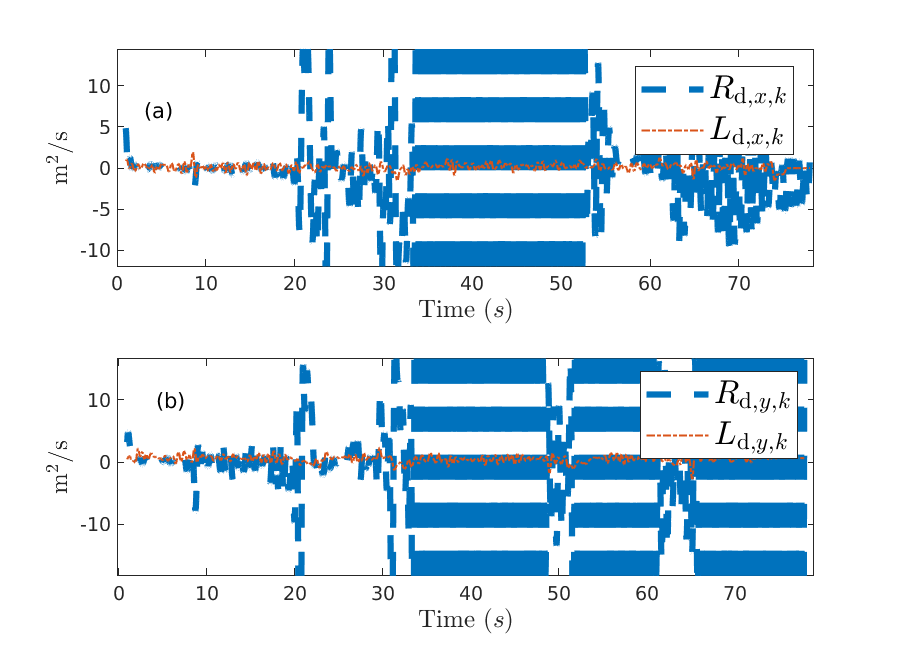}}
            \end{center}
            \caption{ {\it  Example \ref{eg_exp_x_rate_sinu}:}  Rate gyro with sinusoid fault. Beginning at 30 s, $R_{\rmd,{x},k}$ in (a) and $R_{\rmd,{y},k}$ in (b) exhibits sinusoid fault and diverge from $L_{\rmd,{x},k}$ and $L_{\rmd,{y},k}$, respectively.} 
            \label{fig:exp_acc_DT_rate_sinu}
          \end{figure} 
          \begin{figure}[h!t]
              \begin{center}
            {\includegraphics[width=0.7\linewidth]{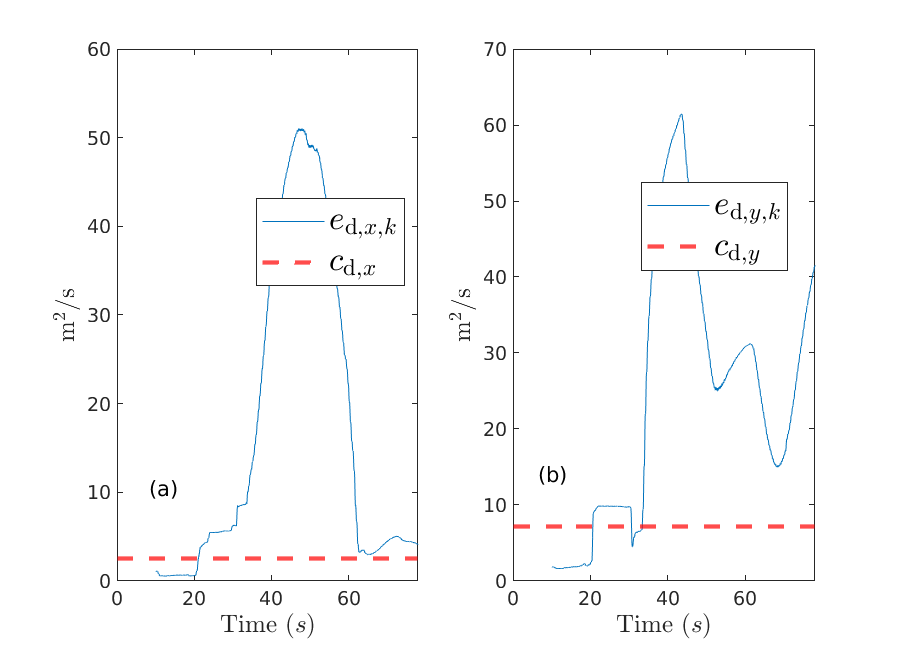}}
            \end{center}
            \caption{{\it  Example \ref{eg_exp_x_rate_sinu}:}  Rate gyro with  sinusoid fault.  $e_{\rmd,{x},k}$ in (a) and $e_{\rmd,{y},k}$ in (b) abruptly increase above the cutoff thresholds $c_{\rmd, x}$ and $c_{\rmd, y}$, respectively, at 30 s when the sinusoid fault begins, which indicates a fault in the sensors used to compute $e_{\rmd,{x},k}$ and $e_{\rmd,{y},k}$. } 
            \label{fig:exp_error_DT_rate_sinu}
          \end{figure} 
          \begin{figure}[h!t]
              \begin{center}
            {\includegraphics[width=0.7\linewidth]{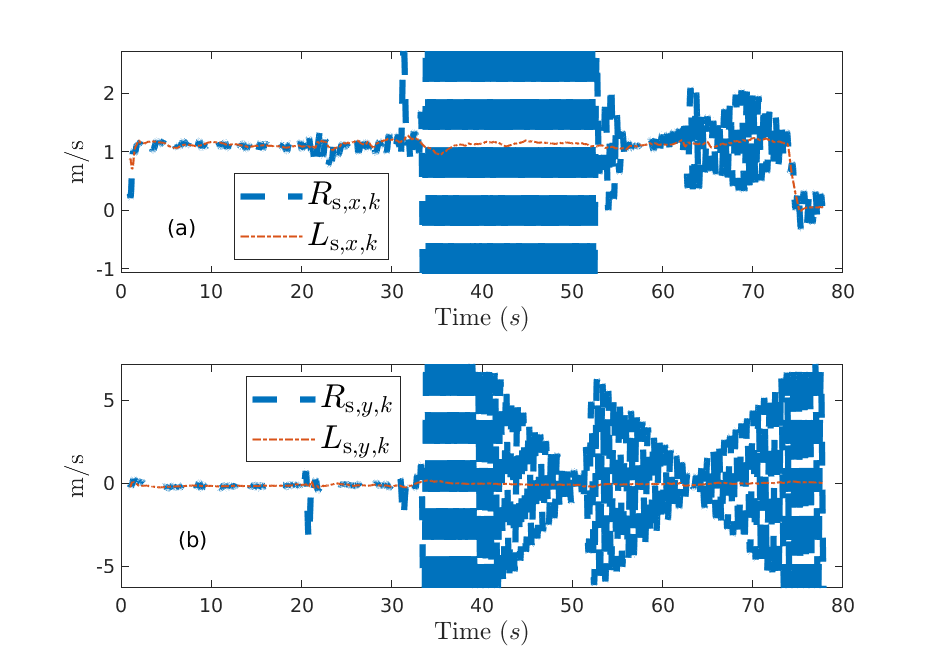}}
            \end{center}
            \caption{{\it  Example \ref{eg_exp_x_rate_sinu}:}  Rate gyro with sinusoid fault. Beginning at 30 s, $R_{\rms,{x},k}$ in (a) and $R_{\rms,{y},k}$ in (b) exhibits sinusoid fault and diverge from $L_{\rms,{x},k}$ and $L_{\rms,{y},k}$, respectively.} 
            \label{fig:exp_acc_ST_rate_sinu}
          \end{figure}
          \begin{figure}[h!t]
              \begin{center}  
              {\includegraphics[width=0.7\linewidth]{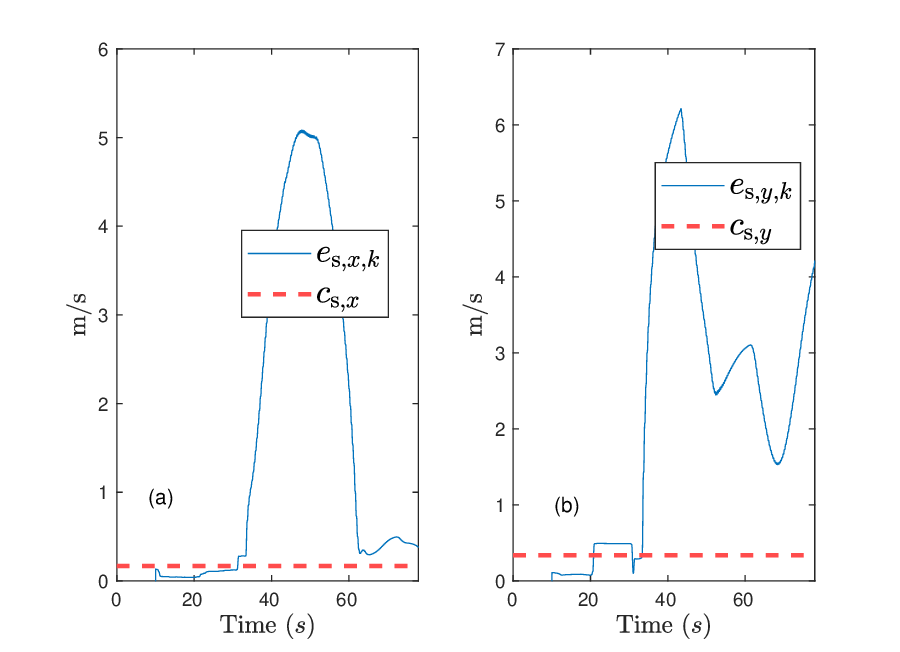}}
            \end{center}
            \caption{{\it  Example \ref{eg_exp_x_rate_sinu}:}  Rate gyro with sinusoid fault. $e_{\rms,{x},k}$ in (a) and $e_{\rms,{y},k}$ in (b) abruptly increase above the cutoff thresholds $c_{\rms, x}$ and $c_{\rms, y}$, respectively, at 30 s when the sinusoid fault begins, exceeding cutoff thresholds, which indicates a fault in the sensors used to compute $e_{\rms,{x},k}$ and $e_{\rms,{y},k}$.} 
            \label{fig:exp_error_ST_rate_sinu}
          \end{figure}
          \begin{figure}[h!t]
              \begin{center}
            {\includegraphics[width=0.7\linewidth]{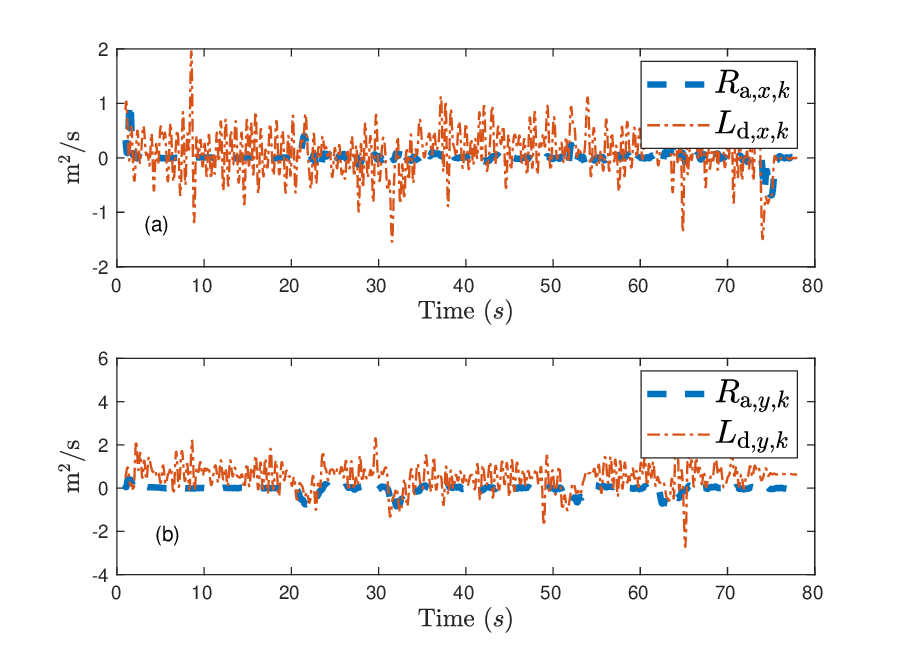}}
            \end{center}
            \caption{{\it  Example \ref{eg_exp_x_rate_sinu}:} Rate gyro with sinusoid fault.  $L_{\rma,{x},k}$ in (a) and $L_{\rma,{y},k}$ in (b) follows $R_{\rma,{x},k}$ and $R_{\rmd,{y},k}$, respectively.} 
            \label{fig:exp_acc_A_rate_sinu}
          \end{figure}
          \begin{figure}[h!t]
              \begin{center}
            {\includegraphics[width=0.7\linewidth]{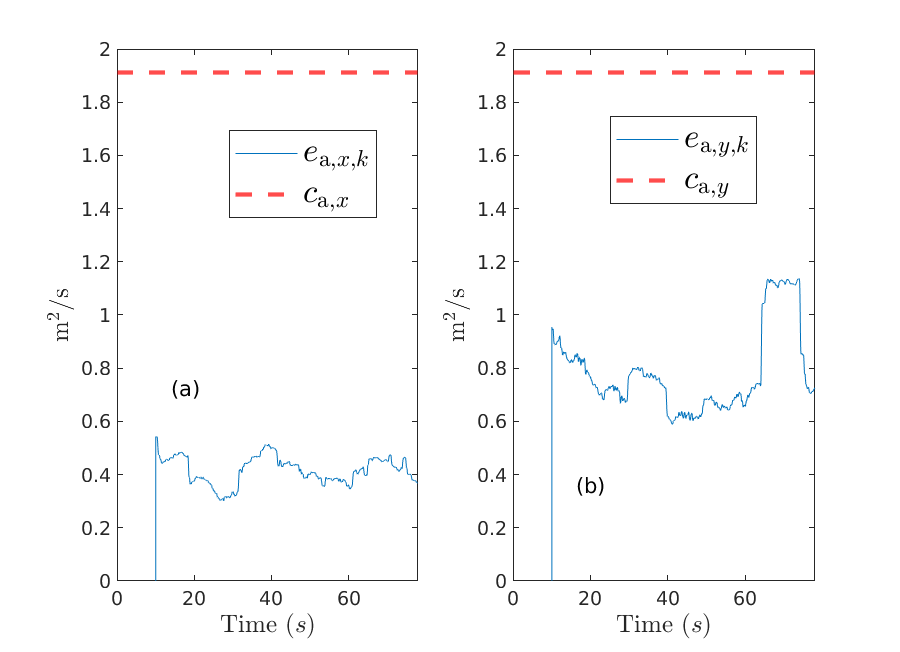}}
            \end{center}
            \caption{{\it  Example \ref{eg_exp_x_rate_sinu}:}  Rate gyro with a sinusoid fault.  $e_{\rma,{x},k}$ and $e_{\rma,{y},k}$ remains below cutoff thresholds.} 
            \label{fig:exp_error_A_rate_sinu}
          \end{figure}
    }      
    \end{exam}

\section{Experimental Application to an Aerial Vehicle}  \label{sec:num_example_exp_3d}

This section provides numerical examples of experimental data to illustrate KSFD in real-world scenarios. This section details the experimental results for a quadcopter in 3D space.

\begin{figure}
     \centering
     \begin{subfigure}[a]{0.6\textwidth}
         \centering
         \includegraphics[width=0.8\linewidth]{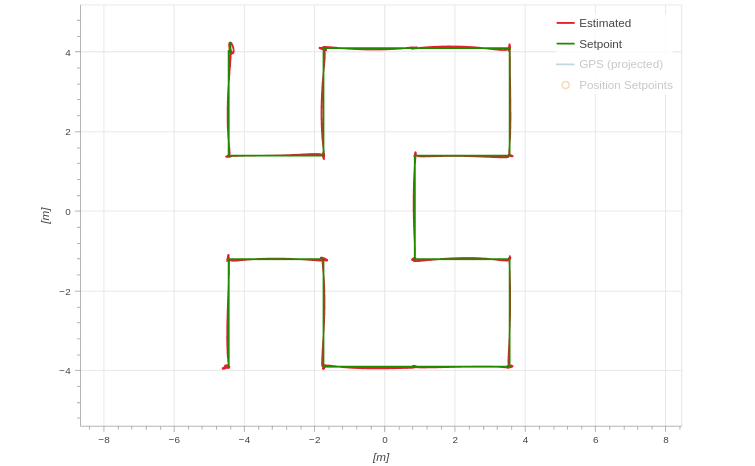}
         \caption{} 
            \label{fig:quad_trajectory}
     \end{subfigure}
     \hfill
     \begin{subfigure}[b]{0.4\textwidth}
         \centering
        \includegraphics[width=0.8\linewidth,angle=-00]{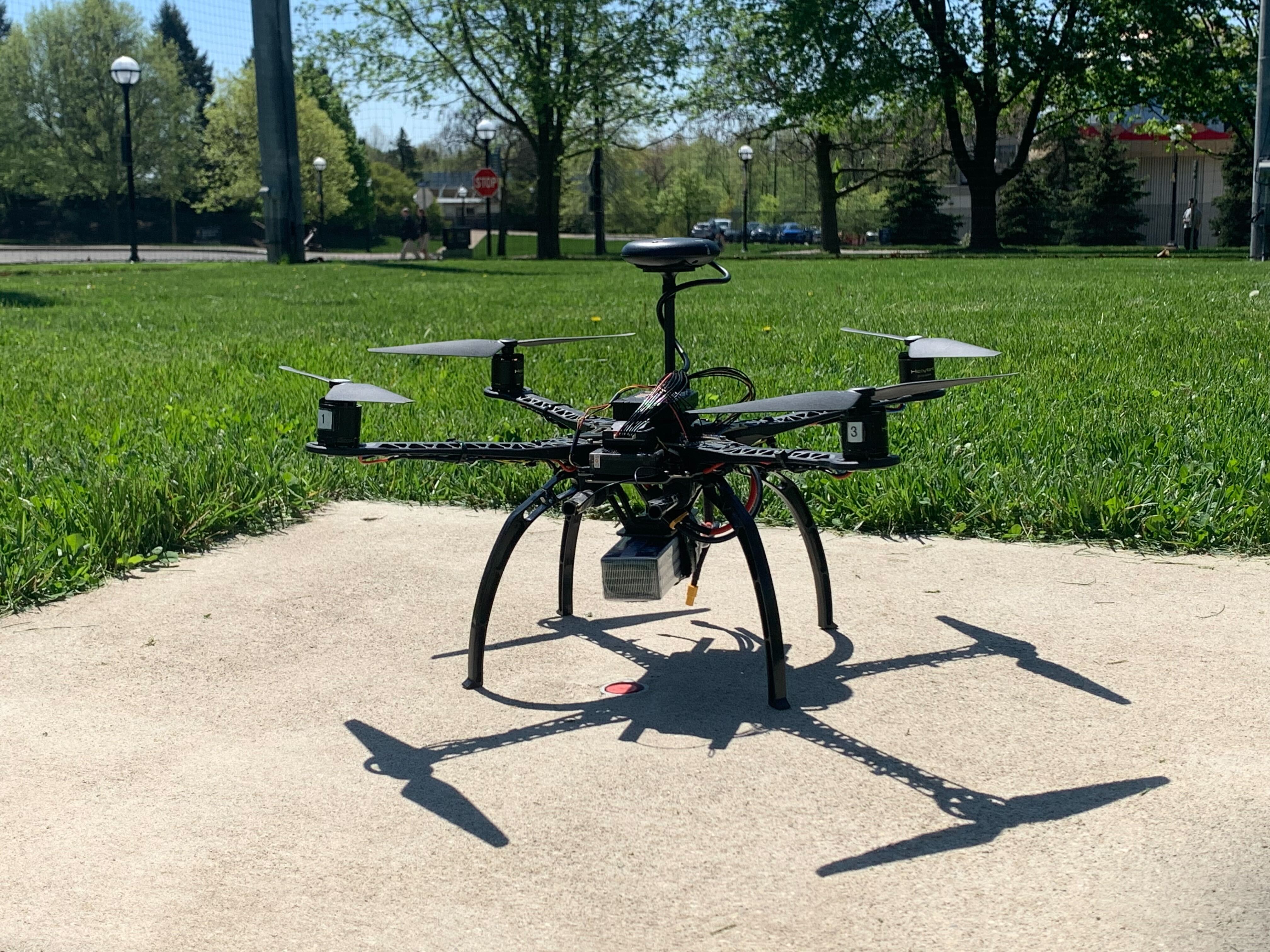}
         \caption{} 
            \label{fig:quad}
     \end{subfigure}
     \caption{(a) Position trajectory of the quadcopter shown in $x-y$ plane. (b) Quadcopter vehicle used in the experiment.}
        % \label{fig:three graphs}
\end{figure}

For the experimental setup, a S500 Quadcopter frame was used, equipped with a Pixhawk 4 autopilot. The quadcopter executed a predefined Hilbert curve, as programmed in the ground control application, as shown in Figure \ref{fig:quad_trajectory}. The sensor data displayed in Table \ref{Tab:SensorData_ground} were obtained from the Pixhawk 4 logs and via a MOCAP system. A $x$-axis accelerometer sensor fault due to bias is analyzed in example \ref{eg_exp_3d_acc_bias}.

In example \ref{eg_exp_3d_acc_bias}, the transient phase for AISE is concluded within $1000$ steps. Consequently, $\delta$ is set at $1000$ steps.
Cutoff thresholds are set at twice the error magnitude at step $k = 3.5\delta$. This parameter choice ensures that the data employed to compute the error metric at step $k = 3.5\delta$ are unaffected by the transient phase and by the fault starting from step $k = 4\delta$.

\begin{exam} \label{eg_exp_3d_acc_bias}
      {\it KSFD for $x$-axis accelerometer failure.}

      {\rm In this scenario, the $x$-axis accelerometer experiences a failure due to a bias of $0.5$ $g$ starting at 40 s. The parameter settings are the same as those used in Example \ref{eg_sim_accelerometer_failure}.

      Figure \ref{fig:exp_acc_DT_acc_bias} compares $R_{\rmd,{x},k}$, $R_{\rmd,{y},k}$, and $R_{\rmd,{z},k}$ with their respective $L_{\rmd,{x},k}$, $L_{\rmd,{y},k}$, and $L_{\rmd,{z},k}$. 
      In Figure \ref{fig:exp_error_DT_acc_bias}, the bias introduced at 40 s leads to a sudden increase in $e_{\rmd,{x},k}$ above the cutoff threshold, which indicates that one of the sensors used to compute $e_{\rmd,{x},k}$ is faulty.
      Figure \ref{fig:exp_acc_ST_acc_bias} displays a comparison between $R_{\rms,{x},k}$, $R_{\rms,{y},k}$, and $R_{\rms,{z},k}$, with $L_{\rms,{x},k}$, $L_{\rms,{y},k}$, and $L_{\rms,{z},k}$. In contrast, Figure \ref{fig:exp_error_ST_acc_bias} indicates that $e_{\rms,{x},k}$, $e_{\rms,{y},k}$, and $e_{\rms,{z},k}$ remain below the cutoff threshold.
        Similarly, Figures \ref{fig:exp_acc_A_acc_bias} and \ref{fig:exp_error_A_acc_bias} highlight faults in the sensors calculating $e_{\rma,{x},k}$.
        Given that $e_{\rmd,{x},k}$ and $e_{\rma,{x},k}$ exceed their respective cutoff thresholds, Table \ref{Tab:Sensor_Diagonastic_ground_vehicle} implies that the $x$-axis accelerometer is faulty.
          \begin{figure}[h!t]
              \begin{center}
            {\includegraphics[width=0.7\linewidth]{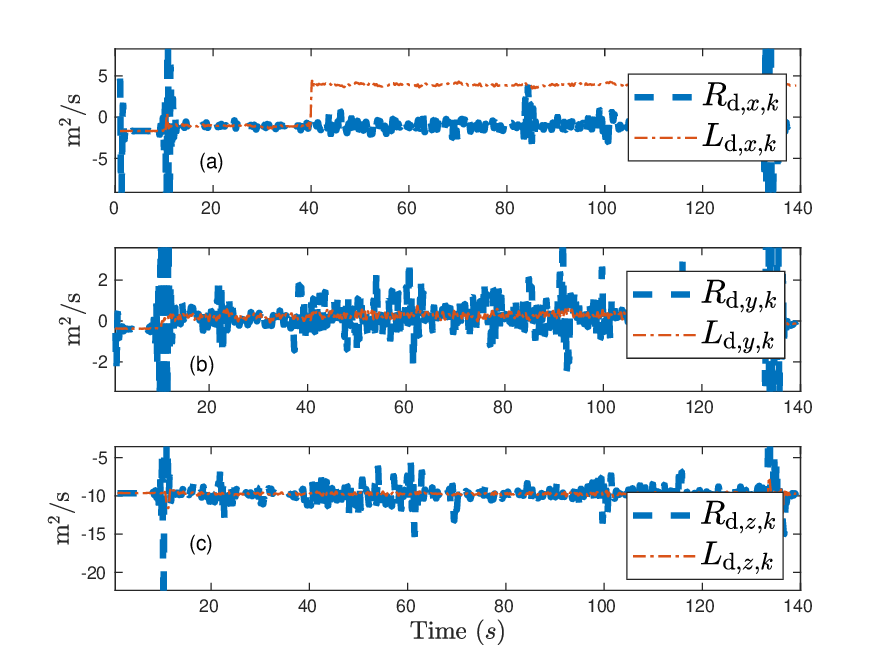}}
            \end{center}
            \caption{ {\it  Example \ref{eg_exp_3d_acc_bias}:}  $x$-axis accelerometer with bias. Beginning at 40 s, $L_{\rmd,{x},k}$ in (a) exhibit bias and thus diverge from $R_{\rmd,{x},k}$. $R_{\rmd,{y},k}$ and $R_{\rmd,{z},k}$ follows $L_{\rmd,{y},k}$ and $L_{\rmd,{z},k}$.}  
            \label{fig:exp_acc_DT_acc_bias}
          \end{figure} 
          \begin{figure}[h!t]
              \begin{center}
            {\includegraphics[width=0.7\linewidth]{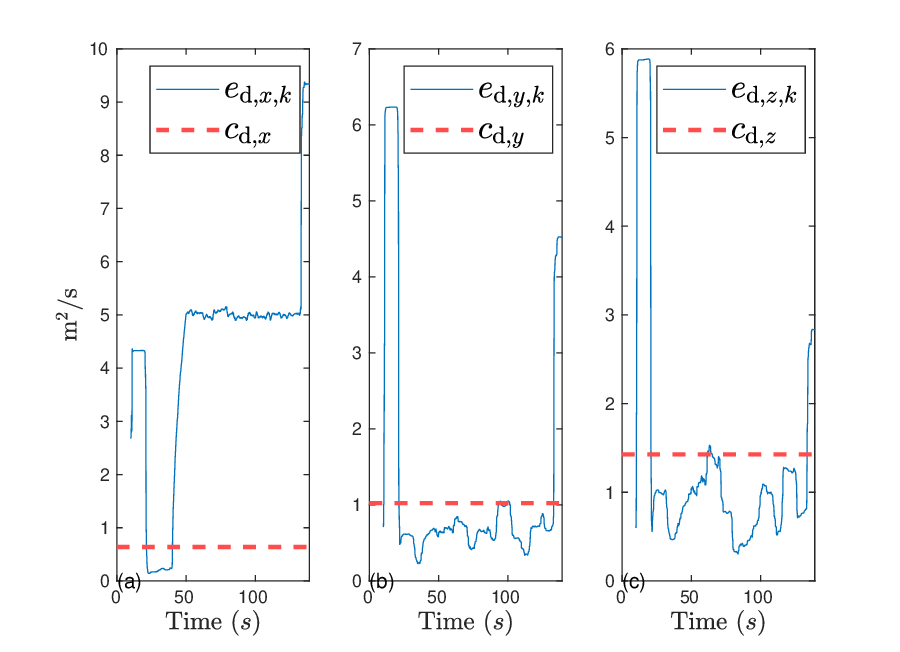}}
            \end{center}
            \caption{ {\it  Example \ref{eg_exp_3d_acc_bias}:}  $x$-axis accelerometer with bias. $e_{\rmd,{x},k}$ in (a) abruptly increases above the cutoff threshold $c_{\rmd,x}$ at 40 s when the bias begins, which indicates that one of the sensors used to compute $e_{\rmd,{x},k}$ is faulty. $e_{\rmd,{y},k}$ and $e_{\rmd,{z},k}$ remain below the cutoff thresholds $c_{\rmd,y}$ and $c_{\rmd,z}$. } 
            \label{fig:exp_error_DT_acc_bias}
          \end{figure} 
          \begin{figure}[h!t]
              \begin{center}
            {\includegraphics[width=0.7\linewidth]{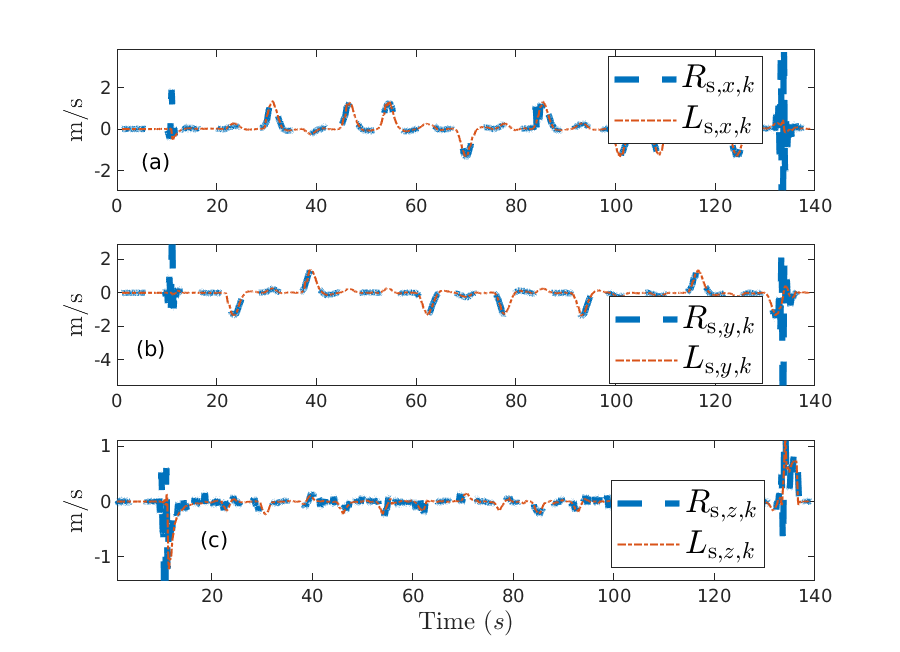}}
            \end{center}
            \caption{ {\it  Example \ref{eg_exp_3d_acc_bias}:}  $x$-axis accelerometer with bias.  $R_{\rms,{x},k}$, $R_{\rms,{y},k}$, and $R_{\rms,{z},k}$ follows $L_{\rmd,{x},k}$, $L_{\rmd,{y},k}$, and $L_{\rmd,{z},k}$ in (a), (b), and (c), respectively.} 
            \label{fig:exp_acc_ST_acc_bias}
          \end{figure}
          \begin{figure}[h!t]
              \begin{center}
            {\includegraphics[width=0.7\linewidth]{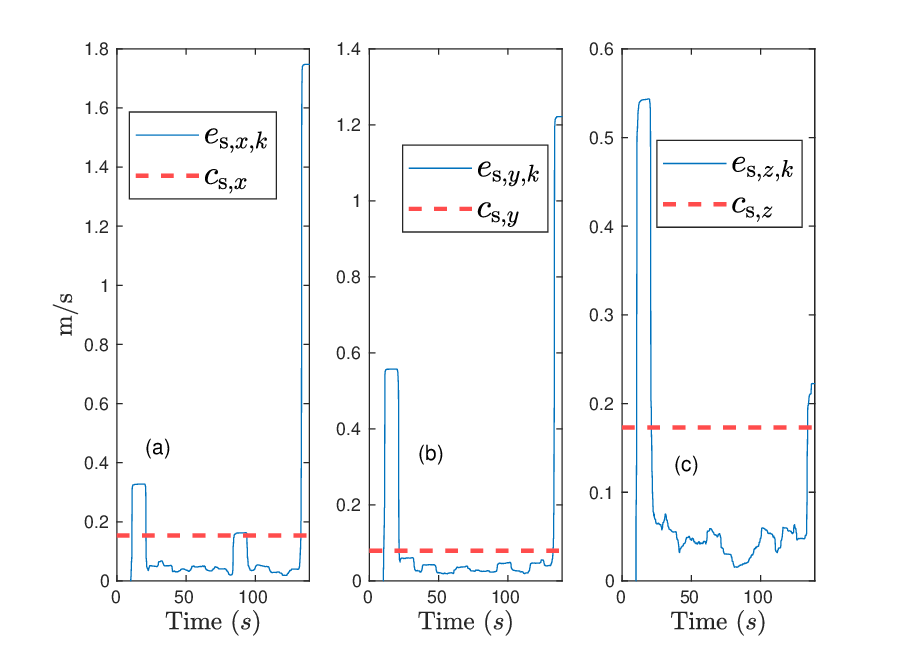}}
            \end{center}
            \caption{ {\it  Example \ref{eg_exp_3d_acc_bias}:}  $x$-axis accelerometer with bias. $e_{\rms,{x},k}$, $e_{\rms,{y},k}$, and $e_{\rms,{z},k}$ remains below the cutoff threshold.} 
            \label{fig:exp_error_ST_acc_bias}
          \end{figure}
          \begin{figure}[h!t]
              \begin{center}
            {\includegraphics[width=0.7\linewidth]{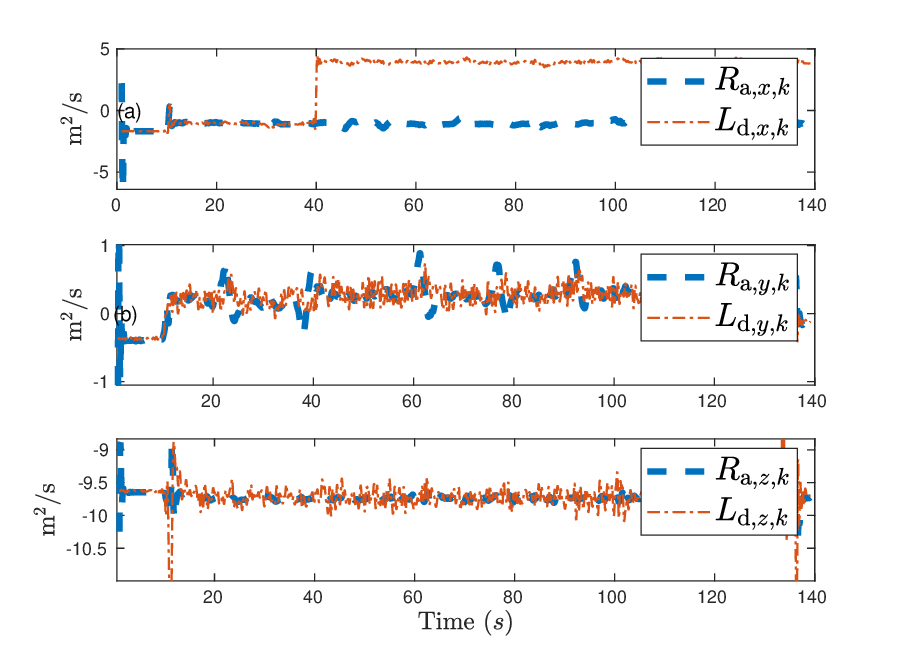}}
            \end{center}
            \caption{ {\it  Example \ref{eg_exp_3d_acc_bias}:}  $x$-axis accelerometer with bias. Beginning at 40 s, $L_{\rma,{x},k}$ in (a) exhibit bias and thus diverge from $R_{\rma,{x},k}$. $R_{\rma,{y},k}$ and $R_{\rma,{z},k}$ follows $L_{\rma,{y},k}$ and $L_{\rma,{z},k}$.} 
            \label{fig:exp_acc_A_acc_bias}
          \end{figure}
          \begin{figure}[h!t]
              \begin{center}
            {\includegraphics[width=0.7\linewidth]{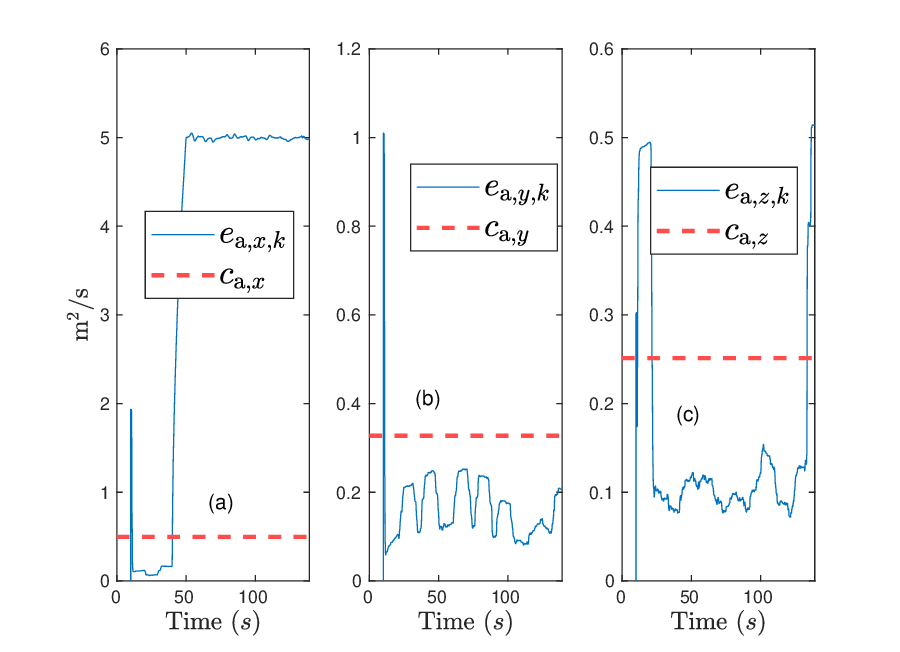}}
            \end{center}
            \caption{ {\it  Example \ref{eg_exp_3d_acc_bias}:}  $x$-axis accelerometer with bias. $e_{\rma,{x},k}$ in (a) abruptly increases above the cutoff threshold $c_{\rma,x}$ at 40 s when the bias begins, which indicates that one of the sensors used to compute $e_{\rma,{x},k}$ is faulty. $e_{\rma,{y},k}$ and $e_{\rma,{z},k}$ remain below the cutoff thresholds $c_{\rma,y}$ and $c_{\rma,z}$. } 
            \label{fig:exp_error_A_acc_bias}
          \end{figure}
    }
    \end{exam}

%%%%%%%%%%%%%%%%%%%%%%%%%%%%%%%%%%%%%%%%%%%%%%%%%%%%%%%%%%%
\section{Conclusions} \label{sec:conclusion}
This paper introduced kinematics-based sensor fault detection (KSFD) for detecting and identifying faulty sensors in autonomous vehicles. KSFD, which is model-independent, leverages kinematic relations, sensor measurements, and real-time single and double numerical differentiation using the adaptive input and state estimation (AISE). By employing kinematics-based error metrics, this approach identifies sensor faults in both ground and aerial vehicles.

For ground vehicles operating in the horizontal plane, KSFD uses six real-time-computed, kinematics-based error metrics from onboard sensor data, including radar, rate gyro, magnetometer, and accelerometer measurements, along with their derivatives. For aerial vehicles, nine kinematics-based error metrics are used to account for the additional degrees of freedom.

KSFD addresses several limitations associated with model-based and knowledge-based methods, such as the need for accurate mathematical models and comprehensive historical data. By relying on exact kinematic relations, real-time sensor data, and AISE, KSFD reduces the complexity and computational burden typically involved in sensor-fault detection.

Simulated and experimental examples demonstrated the effectiveness of KSFD in detecting sensor faults. The results underscore the potential of KSFD for enhancing the reliability and safety of autonomous vehicles by providing robust sensor-fault-detection capabilities.

Future work will focus on refining KSFD to handle simultaneous sensor failures, expanding the applicability to a broader range of vehicles and sensor types, and integrating this technique with vehicle control systems to enable real-time, fault-tolerant operation. 
Finally, more advanced statistical techniques for setting the cutoff thresholds are of interest.

\section*{Acknowledgments}
The authors would like to thank Juan A. Paredes for assisting with the rover and quadrotor experiment data used in this paper. This research was supported by NSF grant CMMI 2031333.



%%       \citet{<label>} ==> Jones et al. [21]
%%       \citep{<label>} ==> [21]
%%


\bibliographystyle{elsarticle-num-names} 
\bibliography{bibpaper,NSFbibfile,sensorfault}




% \end{thebibliography}
\end{document}